\documentclass{aa}
\usepackage{multirow}
\usepackage{txfonts}
\usepackage{graphicx}
\usepackage{natbib}
\usepackage{xcolor, color}
\usepackage{caption}
\usepackage[labelformat=simple]{subfig}
\renewcommand{\thesubfigure}{\relax}
\usepackage{booktabs}
\usepackage{threeparttable}  
\usepackage{comment}
\usepackage{array}

\newcommand{\GG}[1]{}

\begin{document}

   \title{Seeds of Life in Space (SOLIS)}

   \subtitle{ X.
   Interstellar Complex Organic Molecules in the NGC 1333 IRAS 4A outflows}

    \author{M. De Simone\inst{1}
          \and
          C. Codella \inst{2,1}
          \and
          C. Ceccarelli \inst{1}
          \and
          A. L\'opez-Sepulcre \inst{3,1}
          \and
          A. Witzel \inst{1}
          \and
          R. Neri \inst{3}
          \and 
          N. Balucani \inst{4,1}
          \and
          P. Caselli \inst{5}
          \and
          C. Favre \inst{1}
          \and
          F. Fontani \inst{2}
           \and
          B. Lefloch \inst{1}
          \and
          J. Ospina-Zamudio \inst{1}
          \and
          J. E. Pineda \inst{5}
          \and
          V. Taquet \inst{2}  
          }

   \institute{Univ. Grenoble Alpes, CNRS, IPAG, 38000 Grenoble, France
   \and
   INAF, Osservatorio Astrofisico di Arcetri, Largo E. Fermi 5, 50125 Firenze, Italy 
   \and
   Institut de Radioastronomie Millim\'etrique (IRAM), 300 rue de la Piscine, 38400 Saint-Martin d'H\`eres, France
   \and
   Dipartimento di Chimica, Biologia e Biotecnologie, Università degli Studi di Perugia, Perugia 06123, Italy   
   \and
   Max-Planck-Institut f\"ur extraterrestrische Physik (MPE), Giessenbachstrasse 1, 85748 Garching, Germany
}

%

\offprints{M. De Simone, \email{marta.desimone@univ-grenoble-alpes.fr}}

\date{Received: 28/10/2019 ; accepted: 12/06/2020 }

  \abstract
   {The interstellar Complex Organic Molecules (iCOMs) are {C-bearing molecules containing at least six atoms}; two main approaches of their formation are invoked: a direct formation in the icy mantle of the dust grains, or through the reaction in gas phase of released grain mantle species.
   The shocked gas along outflows driven by low-mass protostars is a unique environment to study how the iCOMs can be formed, as the composition of the dust mantles is sputtered into the gas phase.
   }
   {The chemical richness in shocked material associated with low--mass protostellar outflows has been so far studied in the prototypical L1157 blue shifted outflow to investigate the iCOMs formation routes.
   To understand whether the case of L1157-B1 is unique, we imaged and studied the IRAS 4A outflows in the NGC 1333 star forming region.}
   {We used the NOEMA (NOrthern Extended Millimeter Array) interferometer as part of the IRAM SOLIS (Seeds Of Life in Space) Large Program to image the large scale bipolar outflows driven by the IRAS 4A system in the 3 mm band, and we compared the observation with the GRAINOBLE+ astrochemical model. }
   {We report the first detection, in the IRAS 4A outflows, of several iCOMs: six lines of methanol (CH$_3$OH), eight of acetaldehyde (CH$_3$CHO), one of formamide (NH$_2$CHO) and four of dimethyl ether (CH$_3$OCH$_3$), all sampling upper excitation energy up to $\sim$30 K.  
   We found a significant chemical differentiation between the south east outflow driven by the IRAS 4A1 protostar, showing a richer molecular content, and the north--south west one driven by the IRAS 4A2 hot corino. The CH$_3$OH/CH$_3$CHO abundance ratio is lower by a factor $\sim$4 in the former; furthermore the ratio in the IRAS 4A outflows is lower by a factor $\sim$10 with respect to the values found in different hot corinos.} 
   {After L1157-B1, IRAS 4A outflow is now the second outflow to show an evident chemical complexity.
   Given that CH$_3$OH is a grain surface species, the astrochemical gas phase model run with GRAINOBLE+ reproduced our observation assuming that acetaldehyde is formed mainly through the gas phase reaction of ethyl radical (CH$_3$CH$_2$) and atomic oxygen. Furthermore, the chemical differentiation between the two outflows suggests that the IRAS 4A1 outflow is likely younger than the IRAS 4A2 one. 
   Further investigation is needed to constrain the age of the outflow. In addition, observation of even younger shocks are necessary. In order to provide strong constraints on the CH$_3$CHO formation mechanisms it would be interesting to observe CH$_3$CH$_2$, but given that its frequencies are not known, future spectroscopic studies on this species are needed.}

   \keywords{Stars: formation -- ISM: jets and outflows -- 
ISM: molecules -- ISM: individual objects: IRAS 4A}

   \maketitle
%

\section{Introduction}\label{sec:intro}

Since the discovery of interstellar Complex Organic Molecules \citep[iCOMs\footnote{{
Please note that we added “i” to the commonly used COMs acronym in
order to be clear that these molecules are only complex in the interstellar
context, contrary to what chemists would consider complex in the terrestrial
context.}}, molecules containing carbon and at least six atoms:][]{herbst_complex_2009,ceccarelli_seeds_2017} in Solar-type protostars \citep{cazaux_hot_2003}, the question whether they had a role in the appearance of life on Earth (and elsewhere in the Universe) has been raised. Although they are extremely small molecules when compared to the biotic ones, iCOMs may have provided the bricks to build them. The presence of amino acids in meteorites and comets has certainly revived this possibility \citep[e.g.][]{pizzarello_nature_2006,elsila_cometary_2009,altwegg_prebiotic_2016}.

In addition to their possible role in the  {emergence of life}, iCOMs have represented a challenge for astrochemistry, as their synthesis is all but obvious. 
Nowadays, two main paradigms are invoked \citep[see e.g.][]{herbst_synthesis_2017}, that argue that iCOMs are either synthesized on the grain surfaces \citep[e.g.][]{garrod_formation_2006,garrod_new_2008} or in the gas phase \citep[e.g.][]{millar_formation_1991,balucani_formation_2015,skouteris_genealogical_2018}.  {As a starting point}, both pathways have the formation of simple hydrogenated molecules on dust grain mantles during the prestellar phase. 
Constraining which of the two ways to synthesize iCOMs is efficient and where the iCOMs formation happen, is not a simple task. Many methods have been used, from the comparison of the iCOMs measured abundances in hot cores/corinos with model predictions to their measured deuterium fractionation \citep{turner_detection_1990,ceccarelli_detection_1998,coutens_alma-pils_2016, skouteris_new_2017, jorgensen_alma-pils_2018}.  

One method that turned out to be very efficient is to compare observations towards low-mass outflow shocks with model predictions \citep{codella_seeds_2017}. 
{The advantage of this method is that the outflow shocks provide the time dependence as additional constraint. In fact, once localized the iCOMs emission in a precise region (thanks to high spatial resolution observations), it is possible to identify in that region a shock event that corresponds naturally to a precise kinematical age \citep[e.g.][]{gueth_precessing_1996,podio_first_2016}.  
After the passage of the shock, the chemistry in the shocked region evolves with time.
Therefore, the comparison of observed iCOMs abundances with model predictions provides strong constraints on the formation routes because it is possible to do the comparison at the precise kinematical shock age. }
This method was successfully applied in the L1157--B1 outflow shock to constrain the formation route of formamide. In fact, thanks to interferometric high spatial resolution observations, \citet{codella_seeds_2017} found a difference in the spatial distribution between acetaldehyde and formamide emission, and consequently, they could constrain the formamide formation as due to gas-phase reactions. Of course, those conclusions apply to L1157-B1 only. 
Given its power, it is important to apply the same method to other iCOMs and other protostellar shocks.

Unfortunately, observations of iCOMs in low-mass protostellar shocks are very few. To our knowledge, iCOMs other than methanol have been detected only towards a handful of objects: several iCOMs towards L1157-B1 \citep{arce_complex_2008,lefloch_l1157-b1_2017}, 
formamide towards L1157-B2 \citep{mendoza_molecules_2014}, acetaldehyde towards IRAS 2A and IRAS 4A \citep{holdship_observations_2019}  {and acetaldehyde and dimethyl ether towards SMM4-W \citep{oberg_complex_2011}}. However, it is worth noticing that all these works refer to (relatively low spatial angular resolution) single-dish observations and are, by definition, unable to disentangle the different spatial distribution of iCOMs caused by the age of the shocks, so that the method described above cannot be used.

In this work, we present new high spatial observations towards the two outflows from IRAS 4A.
This source is one of the target of the Large Program SOLIS \citep[Seeds Of Life In Space:][]{ceccarelli_seeds_2017}, at the IRAM/NOEMA (NOrthern Extended Millimeter Array) interferometer, whose goal is to investigate the iCOM chemistry during the earliest formation phases of  {Solar--type stellar systems}. 
The observations targeted three iCOMs in addition to methanol (CH$_3$OH): acetaldehyde (CH$_3$CHO), dimethyl ether (CH$_3$OCH$_3$) and formamide (NH$_2$CHO). All these iCOMs were detected in our data set. 
{The detection of different iCOMs in the outflowing gas of IRAS 4A with high spatial resolution observations allows us to apply the method of model-observations comparison described above.}

The article is organized as follows: 
we first give the IRAS 4A source background in \S \ref{sec:iras4a}, then present the observations in \S \ref{sec:observ} and the results in \S \ref{sec:results}; we derive the abundance ratios of the detected iCOMs in different positions of the IRAS 4A outflows (\S \ref{sec:CDratios}), and the model predictions to interpret them (\S \ref{sec:modelling}); in \S \ref{sec:discussion} we discuss what our new observations imply and finally, \S \ref{sec:conclusions} summarizes our work.

\section{IRAS 4A: Source Background}\label{sec:iras4a}
IRAS 4A is part of the multiple system IRAS 4, located at a distance of 299$\pm$15 pc in the NGC 1333 region of the Perseus complex \citep[][]{zucker_mapping_2018}. The system IRAS 4A is constituted by four objects: 4A, 4B, 4B${'}$ and 4C \citep{lay_ngc_1995,looney_unveiling_2000, smith_ngc_2000, di_francesco_infall_2001, choi_high-resolution_2001}. 
IRAS 4A is itself a binary system with two Class 0 objects, 4A1 and 4A2, separated by 1$\farcs$8 \citep[$\sim$ 540 au;][]{looney_unveiling_2000,santangelo_jet_2015,lopez-sepulcre_complex_2017,maury_characterizing_2019}. In the millimeter wavelengths, 4A1 is three times brighter than 4A2. However, their respective luminosity is unknown since they are not resolved in the sub-millimeter to IR wavelengths where the luminosity peak lies. 
The bolometric luminosity of the whole IRAS 4A system is 9.1 L$_\odot$ \citep{kristensen_water_2012,karska_water_2013}.

IRAS 4A is the second ever discovered hot corino \citep{bottinelli_complex_2004}, after IRAS 16293-2422 \citep{cazaux_hot_2003}. Interferometric IRAM/PdBI (Plateau de Bure Interferometer, now evolved into NOEMA) observations have later suggested that iCOMs emission originates rather from 4A2 than 4A1 \citep{taquet_constraining_2015,de_simone_glycolaldehyde_2017}. More recently, \citet{lopez-sepulcre_complex_2017} obtained high resolution ($\sim$0$\farcs$5) ALMA images of IRAS 4A and confirmed the huge contrast between 4A1 and 4A2: while 4A2 shows a hot corino activity with enriched iCOMs emission, no sign of iCOMs is detected in 4A1. 
\citet{lopez-sepulcre_complex_2017} suggest that either 4A1 does not host any hot corino or, alternatively, the hot corino size is less than $\sim$15 au \citep[after scaling to 299 pc the distance adopted by][]{lopez-sepulcre_complex_2017}, namely six times smaller than the 4A2 one.

As for many Class 0 protostars, IRAS 4A is associated with a spectacular large--scale (few arcminutes) bipolar outflow observed with several tracers, such as CO, SiO, SO, HCN \citep{blake_molecular_1995,lefloch_cores_1998,choi_variability_2005,choi_radio_2011}.  \citet{choi_variability_2005} well traced the high velocity component with SiO emission using VLA observations at 2$''$ spatial resolution. From  {their} map, it is possible to distinguish two different blue--shifted lobes towards south and only one northern red--shifted lobe with a peculiar bending toward north--east at 20$''$ from the protostars. 
Using IRAM/PdBI high spatial resolution observations ($<1''$), \citet{santangelo_jet_2015} mapped the outflows at lower scale ($\sim$30$''$) with respect to the SiO map from \citet{choi_variability_2005}. They traced different velocity component (from $\sim$10 km s $^{-1}$ to $\sim$ 60 km s $^{-1}$) using CO, SiO and SO as tracers. 
With their study, \citet{santangelo_jet_2015} were able to clearly disentangle the two southern lobes revealing a fast collimated jet associated with bright H$_2$ emission and driven by 4A1 (south--east lobe) and a slower and precessing jet driven by 4A2 (south--west lobe). 
Furthermore, the jets present different morphologies: 
the 4A2 jet shows a large spatial extent and a S--shape pattern on small scales, probably due to jet precession;
the 4A1 jet is faster than the 4A2 one, covers a smaller extent ($\sim$ 15$''$) and presents a C-shape tilted towards east of 4A1. 

Thanks to a detailed study on sulfur species using interferometric observations, for the first time, \citet{taquet_seeds_2019} were able to distinguish the outflow driven by 4A1 from the one driven by 4A2 also in the northern lobe.

\begin{figure*}
    \centering
    \includegraphics[scale=0.23]{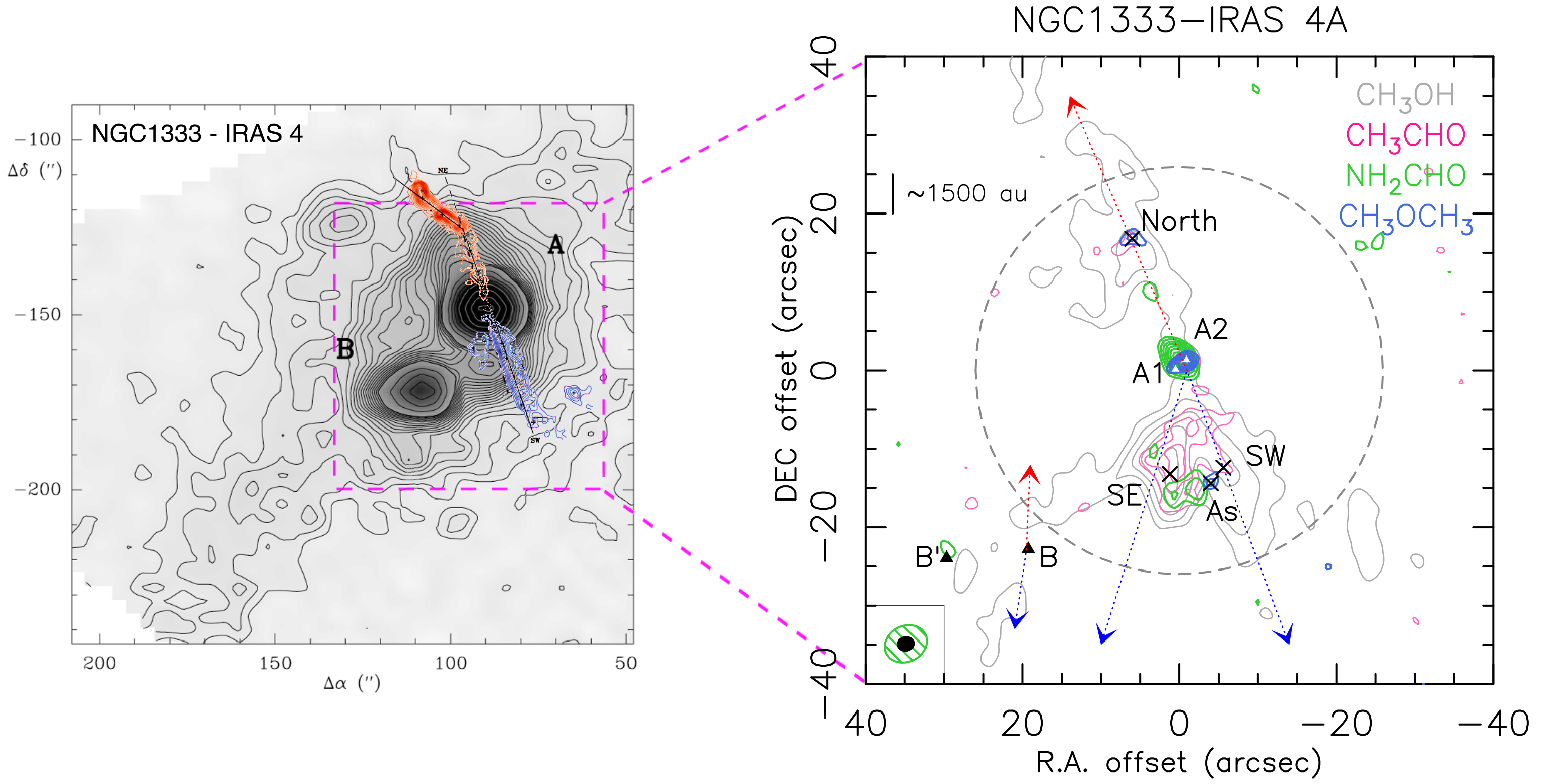}
    \caption{\textit{Left:} {Overlap of the contour map of the 1.25 mm continuum emission from NGC 1333 IRAS 4 region in Perseus, observed with the IRAM 30m antenna \citep{lefloch_cores_1998}, with the map of SiO line \citep[VLA observations;][]{choi_variability_2005}. Axes offsets are in arcseconds from SVS 13 ($\rm \alpha(2000)=03^h 29^m 3\fs9$ and $\delta(2000) =31^\circ 16' 8'')$.
    \textit{Right: }Zoom in in the IRAS 4A system with NOEMA-SOLIS observations. Axes offsets are in arcseconds from IRAS 4A. A spatial separation of 5$''$ correspond to $\sim 1500$ au at a distance of 299 pc \citep{zucker_mapping_2018}. }
    The white triangles mark the position of the sources 4A1 and 4A2, while the black triangles mark the position of the sources 4B and 4B$'$ (coordinates in Table \ref{tab:coordinates}). The black crosses mark the analyzed positions in the outflows (SE, SW, North and As; coordinates in Table \ref{tab:coordinates}).  The dashed blue and red arrows indicate the direction of the blue-- and red--shifted 4B outflow \citep[from the HCN observations of][]{choi_high-resolution_2001} and of the 4A ones. 
    The contour map represents the iCOMs emission at 3 mm in the IRAS 4A outflows (this work). 
    For all the iCOMs the contours start at 3$\sigma$ with steps of 1$\sigma$, except for methanol whose contours have steps of 20$\sigma$. 
    {The emission distribution is the following:  1) methanol (CH$_3$OH in grey), integrated over the transitions $2_{0,2}-1_{0,1}$ A, $2_{0,2}-1_{0,1}$ E and $2_{-1,2}-1_{-1,1}$ E  with $\sigma=75$ mJy/beam km/s; 2) acetaldehyde (CH$_3$CHO, in magenta), here in the $5_{0,5}-4_{0,4}$ A emission with $\sigma=11 $mJy/beam km/s; 3) formamide (NH$_2$CHO, in green), $4_{1,4}-3_{1,3}$ emission with $\sigma=10$ mJy/beam km/s; dimethyl ether (CH$_3$OCH$_3$, in blue), $4_{1,4}-3_{0,3}$ emission with $\sigma=9$ mJy/beam km/s. 
    The synthesized beams for the formamide line (green, $\sim 4''$) and for the other species (black, $\sim 2''$) are indicated in the lower left corner. The primary beam ($\sim 52''$) is shown with a dashed grey circle.}
    }
    \label{fig:Contour map}
\end{figure*}

The left panel of figure \ref{fig:Contour map} summarizes the situation: it shows the distribution of the dust cores at large scale, trace by the continuum at 1.3 mm using the IRAM 30m  \citep{lefloch_cores_1998}, together with large scale high velocity outflow traced by the SiO (1-0) line using the VLA interferometer\citep{choi_variability_2005}.

\section{Observations}\label{sec:observ}
\begin{table*}
    \centering
    \caption{List of the characteristics of the SOLIS WideX backend setups.}
    \label{tab:obs_infos}
    \begin{threeparttable}
    \begin{tabular}{c|ccccccccc}
    \hline
    \hline
     Setup    & Frequency Range (GHZ) & \multicolumn{2}{c}{Spectral resolution}  & \multicolumn{2}{c}{Spatial resolution} & Synthesized Beam & \multicolumn{2}{c}{Primary Beam} \\
         & [GHz]    &   [km s$^{-1}$] & [MHz] & $['']$ & [au]$^a$ & [$''\times ''$ ($\degr$)] & $['']$& [au]$^a$\\
    \hline
    1 &   80.8-84.4   &   7 & 2 & 4 & $\sim$1200 & 4.5$\times$3.5 (27) & 61$\farcs$4 & $\sim$2$\times$10$^4$\\ 
     3 &   95.5-99.5   &   6  & 2 & 4 & $\sim$1200 & 2.2$\times$1.9 (96) & 59$\farcs$2 & $\sim$2$\times$10$^4$\\
     \hline
    \end{tabular}
    \begin{tablenotes}
    \item[a] computed at the distance of the NGC 1333 region \citep[$\sim$299 pc;][]{zucker_mapping_2018}.
    \end{tablenotes}
    \end{threeparttable}
\end{table*}
IRAS 4A was observed with the IRAM/NOEMA interferometer during several tracks in June and September 2016. 
Two frequency setups were used, called 1 and 3 in \citet[][Table 4]{ceccarelli_seeds_2017}, centered at $\sim$82 and $\sim$97 GHz, respectively.
The array was used in configurations D and C with baselines from 15 m to 304 m for Setup 3 and from 16 m to 240 m for Setup 1.
Here, we present the data obtained using the WideX backend, whose characteristics are summarized in Table \ref{tab:obs_infos}

The phase center is on the IRAS 4A1 source, whose coordinates are listed in Table \ref{tab:coordinates}.
The bandpass was calibrated on 3C454.3 and 3C84, while the flux was calibrated using MWC349 and LKHA101. The calibration of phase and amplitude was done observing 0333+321.
The system temperatures ranged typically between 50 and 200 K.
{The calibration error associated to the absolute flux is $\leq$ 15\%.}
The data were reduced using the packages CLIC and MAPPING of the GILDAS\footnote{http://www.iram.fr/IRAMFR/GILDAS} software collection. 
The data were self-calibrated in phase only; the self-calibration solutions were applied to the data spectral cube, which was then cleaned.
A continuum map (see Figure \ref{fig:continuum maps}) was obtained by averaging line-free channels from the self--calibrated data. 
The resulting synthesized beam is 2$\farcs$2$\times$1$\farcs$9 (P.A.=96$\degr$), for Setup 3, and 4$\farcs$5$\times$3$\farcs$5 (P.A.=27$\degr$) for Setup 1. 
The half power primary beam is 59$\farcs$2 and 61$\farcs$4 for Setup 3 and Setup 1 respectively. 

\section{Results}\label{sec:results}

\subsection{Dust continuum emission}
\begin{figure}
    \centering
    \includegraphics[scale=0.35]{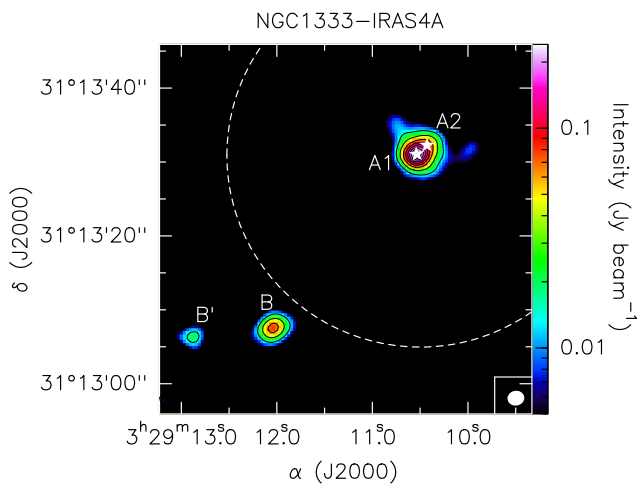}
    \caption{Dust continuum emission maps of IRAS 4A at 95.85-99.45 GHz (Setup 3, see Table \ref{tab:obs_infos}). Contours start at 3$\sigma$ and increase by steps of 20$\sigma$, with $\sigma=1.5$mJy/beam. The synthesized beam ($\sim 2''$) is represented in white in the lower right corner of the panel, the primary beam ($\sim 52''$) is shown with a dashed white circle. {The millimeter continuum sources in the field are labeled following the nomenclature used by \citet{choi_high-resolution_2001} for 4A1 and 4A2 and \citet{di_francesco_infall_2001} for B and B$'$.}}
    \label{fig:continuum maps}
\end{figure}

\begin{table}
    \centering
    \caption{Coordinates of the protostars, \citep[see also:]{choi_high-resolution_2001,di_francesco_infall_2001,lopez-sepulcre_complex_2017,maury_characterizing_2019}, and the analyzed emission peaks (chosen from methanol and dimethyl ether emission, see text) in the outflows.}
    \label{tab:coordinates}
    \begin{tabular}{l|ccccc}
    \hline
    \hline
     Position  & $\alpha$(2000) & $\delta$(2000) \\ 
     \hline
     4A1 & ${\rm 03^h29^m10\fs536}$ & 31$^\circ$13$'$31$\farcs$07\\
     4A2 & ${\rm 03^h29^m10\fs428}$ & 31$^\circ$13$'$32$\farcs$27\\
     4B & ${\rm 03^h29^m12\fs000}$ & 31$^\circ$13$'$08$\farcs$10\\
     4B$'$ & ${\rm 03^h29^m12\fs813}$ & 31$^\circ$13$'$06$\farcs$97\\
     \hline
     SE peak & ${\rm 03^h29^m10\fs591}$ & 31$^\circ$13$'$17$\farcs$53 \\
     SW peak & ${\rm 03^h29^m10\fs061}$ & 31$^\circ$13$'$18$\farcs$61 \\
     North peak & ${\rm 03^h29^m10\fs966}$ & 31$^\circ$13$'$47$\farcs$87 \\
     As region & ${\rm 03^h29^m10\fs184}$ & 31$^\circ$13$'$16$\farcs$62 \\
     \hline
    \end{tabular}
\end{table}

Figure \ref{fig:continuum maps} shows the map of the dust continuum emission at 3 mm, whose emission peaks at the position of 4A1 and 4A2. 
As expected, the two sources are not disentangled as the angular resolution of $\sim2''$ is too close to their angular separation ($\sim 1\farcs8$; \S \ref{sec:observ}).  
In addition, the two protostars IRAS 4B and IRAS 4B$'$ \citep[e.g. ][]{looney_unveiling_2000,choi_high-resolution_2001,di_francesco_infall_2001, maury_characterizing_2019} were detected, even if they were located outside the primary beam of our observations ($\sim 52 ''$). 
The coordinates of all the four protostars are reported in Table \ref{tab:coordinates}.

The root mean square (RMS) noise level is 1.5 mJy/beam and the peak flux towards IRAS 4A1+4A2 is 240$\pm$40 mJy/beam. Taking the error on the measured flux into account and considering the slightly different wavelength (2.7 mm) and angular resolution ($\sim 1\farcs2$) of the observations, this flux value is consistent with the one found by \citet{lopez-sepulcre_complex_2017}. The uncertainties in the flux measurements include the amplitude calibration error ($\sim 15 \%$) that dominates the RMS. 

\subsection{Line emission: maps}
The present observations allow us to image both 4A1 and 4A2 and their molecular outflows. The study of the molecular content around 4A1 and 4A2 protostars is out of the scope of the present paper. Instead, we focus here on the molecular composition of the outflows.

Several lines from methanol (CH$_3$OH), acetaldehyde (CH$_3$CHO),  dimethyl ether (CH$_3$OCH$_3$) and formamide (NH$_2$CHO) were detected along the outflows with a signal to noise ratio (S/N) larger than 3. Table \ref{tab:fit_results} lists the detected lines with their spectroscopic properties.
In Setup 3, we detected six lines of methanol which cover a range of upper level energy (E$_{\rm up}$) from 7 to 28 K, eight lines of acetaldehyde with E$_{\rm up}$ between 13 and 23 K, and four lines of dimethyl ether, blended together and all with E$_{\rm up}$=10 K.
In Setup 1, we detected one line of formamide with E$_{\rm up}$ of 28 K. 

Figure \ref{fig:Contour map} shows the distribution of the line emission of the four detected iCOMs. 
To obtain the methanol map we integrated from -36 km s$^{-1}$ to 36 km s$^{-1}$ with respect to the systematic velocity of the source ($\sim$ 6.5 km s$^{-1}$), for acetaldehyde between -15 km s$^{-1}$ to 15 km s$^{-1}$, while for formamide and dimethyl ether we integrated from -9 km s$^{-1}$ to 9 km s$^{-1}$. 
First, the methanol emission is extended ($\sim 1'$) and covers the lobes of the two outflows from the two protostars: north (North) and south-west (SW) lobes of the outflow from 4A2, and the south-east (SE) lobe from 4A1. Second, acetaldehyde emission is less extended than the methanol one($\sim 15''$) and it is bright towards the southern lobes, especially towards the SE one. 
The dimethyl ether emission is not resolved being less than the beam size ($2 ''$) in two positions, in the North lobe and in the region As \citep[named by][]{ceccarelli_seeds_2017} where SE and SW lobes seem to cross. Finally, formamide emission is also compact ($\sim 6 ''$) and is located around the As position \citep[see also][]{ceccarelli_seeds_2017}. 
{The same iCOMs are also detected in the central protostars (4A1+4A2): please note that the methanol and acetaldehyde emission is not visible in Figure \ref{fig:Contour map} because hidden by the dimethyl ether and formamide contours.}
{Figure \ref{fig:Contour map} clearly shows a first important result: the evidence of a spatial segregation between the different iCOMs due to the fact that their emission covers different outflow regions.}

\subsection{Line emission: spectra and intensities}
\begin{figure*}
    \centering
   \includegraphics[scale=0.6]{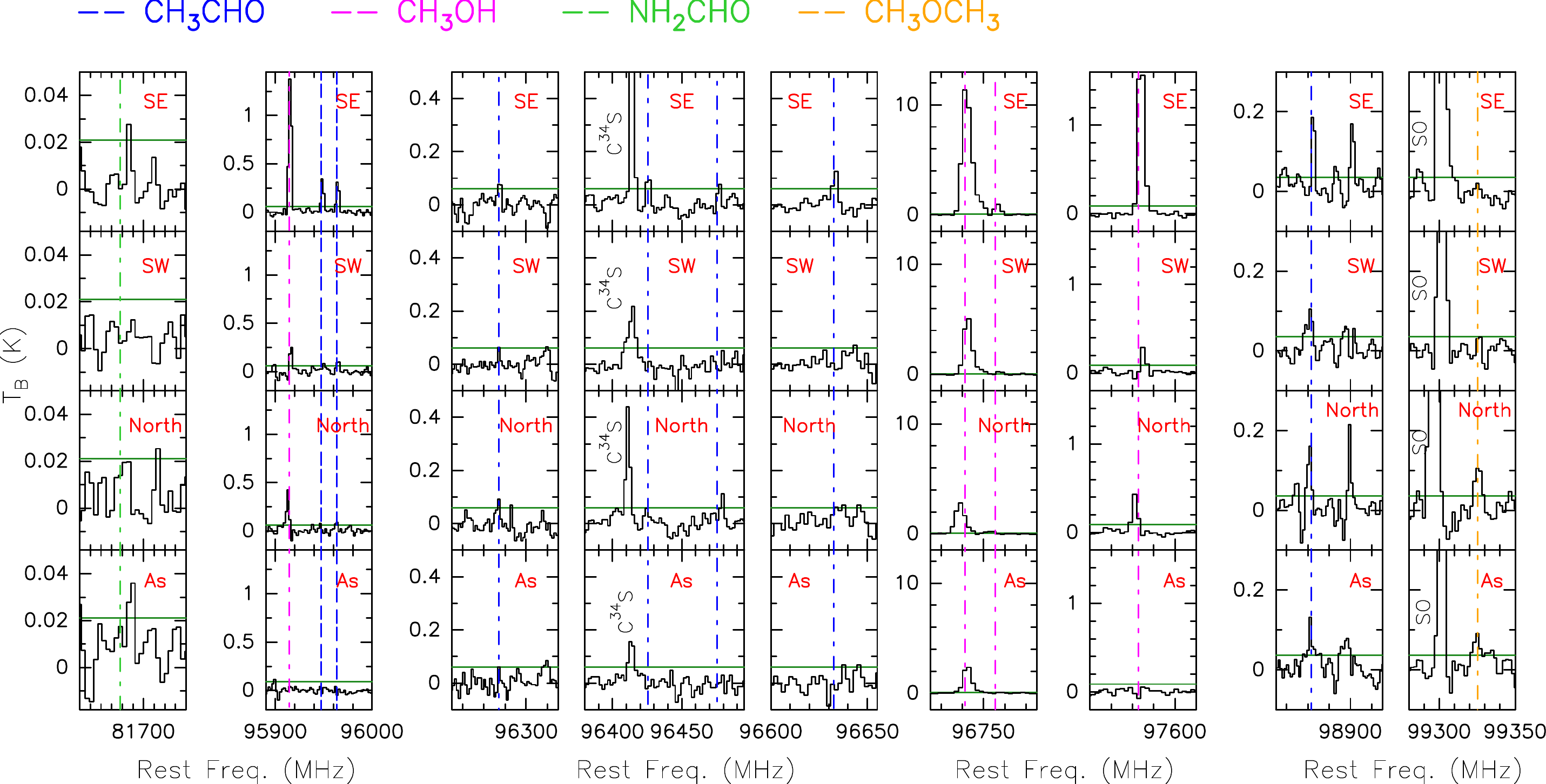}
    \caption{{Spectra towards the four positions along the outflows of IRAS 4A, listed in Table \ref{tab:coordinates}. The horizontal green lines represent the 3$\sigma$ levels (reported in Table \ref{tab:fit_results}); the vertical dashed lines in blue and magenta represent the rest frequency of acetaldehyde and methanol lines, respectively. The rest frequency corresponds to the protostar LSR velocity ($6.5$ km s$^{-1}$).} }
    \label{fig:Spectra_TOT}
\end{figure*}

In order to do a quantitative analysis, we extracted the spectra from different positions of the three lobes, where both methanol and acetaldehyde emit. The first two selected positions correspond to the emission peaks of methanol in the two southern lobes, named SE and SW, while the last selected position corresponds to the emission peak of dimethyl ether in the north lobe, named North (Figure \ref{fig:Contour map}). A fourth position is the one where formamide and dimethyl ether emit, named As.  {The spectra extracted at the pixels corresponding to the four positions are shown in Fig. \ref{fig:Spectra_TOT}}. It is immediately evident that the SE position is richer and brighter in iCOMs when compared to the other ones (SW, North and As). 

We then derived the velocity-integrated line intensities of each detected transition using a Gaussian fit, obtained with the CLASS package of the GILDAS software. All the lines used for the analysis are not contaminated by other species and are well isolated. 
For instance, three of the six detected methanol lines, namely 2$_{\rm -1,2}$--1$_{\rm -1,1}$ E (E$_{\rm up}$=13 K), 2$_{\rm 0,2}$--1$_{\rm 0,1}$ A (E$_{\rm up}$=7 K), 2$_{\rm 0,2}$--1$_{\rm 0,1}$ E (E$_{\rm up}$=21 K), are blended together at the WideX resolution ($\sim$2 MHz): therefore they were not used in the analysis described in the next section.
Table \ref{tab:fit_results} reports the fit results; in case of non-detection, is reported the 3$\sigma$ limit.

\begin{table*}

    \centering
    \caption{{Spectral parameters and fit results of the detected iCOMs emission lines observed using the NOEMA WideX backend towards the IRAS 4A outflow peaks (see text and Table \ref{tab:coordinates}).}}
    \label{tab:fit_results}
	\resizebox{\textwidth}{!}{%
    \begin{threeparttable}
    \begin{tabular}{l|ccc|ccc|ccc|ccc|ccc}
    \hline
    \hline
    \multicolumn{1}{c|}{Transition}&
    \multicolumn{3}{c|}{Spectral Parameters}&
    \multicolumn{3}{c|}{Outflow SE} & 
    \multicolumn{3}{c|}{Outflow SW} &
    \multicolumn{3}{c}{Outflow North} &
    \multicolumn{3}{c}{Region As $^{(a)}$} \\
    \hline
       & Frequency$^{(b)}$	& E$_{\rm up}^{(b)}$  & logA$_{\rm ij}^{(b)}$ & Area$^{(c)}$ & T$_{\rm peak}$ & RMS$^{(d)}$ & Area$^{(c)}$ & T$_{\rm peak}$ & RMS$^{(d)}$ & Area$^{(c)}$& T$_{\rm peak}$ & RMS$^{(d)}$  & Area$^{(c)}$& T$_{\rm peak}$ & RMS$^{(d)}$  \\
     & [GHz] & [K] &  & [K km s$^{\rm -1}$] & [K] & [mK] & [K km s$^{\rm -1}$] & [K] & [mK] & [K km s$^{\rm -1}$] & [K] & [mK] & [K km s$^{\rm -1}$] & [K] & [mK]   \\
    \hline
    
    \multicolumn{16}{c}{CH$_3$OH} \\
    \hline
    2$_{\rm 1,2}$--1$_{\rm 1,1}$ A      & 95.91431	& 	21.4	&	-5.6 &
     17.1(0.9) & 1.4 & 20 &
    2.7(0.3) & 0.4 & 20 & 
    5.1(0.5) & 0.4 & 20 &
    $\leq$ 0.5 & $\leq$ 0.05 & 30  \\
    2$_{\rm -1,2}$--1$_{\rm -1,1}$ E $^{(e)}$    & 		96.73936	& 	12.5	&	-4.6 & 
     \multirow{3}{*}{199 (4)} & \multirow{3}{*}{12} & \multirow{3}{*}{20}  &  
    \multirow{3}{*}{69(4)} & \multirow{3}{*}{5} & \multirow{3}{*}{30}  &
    \multirow{3}{*}{48(4)} & \multirow{3}{*}{3} & \multirow{3}{*}{20} &
    \multirow{3}{*}{37(1)} & \multirow{3}{*}{2.5} & \multirow{3}{*}{30}  \\   
    2$_{\rm 0,2}$--1$_{\rm 0,1}$ A $^{(e)}$ & 96.74138 &  6.9 & -5.6  &     
    &  &  &   
    &  &  &
    &  &   \\   
    2$_{\rm 0,2}$--1$_{\rm 0,1}$ E  $^{(e)}$  &   96.74454	& 	20.1	&	-5.5 &
    &  & &   
    &  &  & 
    &  &   \\
    2$_{\rm 1,1}$--1$_{\rm 1,0}$ E      &  96.75550  	& 	28.0	&	-5.5 &
    11.2(0.3) & 0.9 & 20 & 
    2.0(0.2) & 0.3 & 30 &
    3.2(0.3) & 0.2 & 20 \\   
    2$_{\rm 1,1}$--1$_{\rm 1,0}$ A     & 97.58280  	&	21.6	&	-5.6 & 
    21.5(0.2) & 1.9 & 20 &  
    2.5(0.3) & 0.3 & 30 &
    4.4(0.3) & 0.4 & 30 &
   $\leq$ 0.5 & $\leq$ 0.05 & 30\\
    \hline    
    \multicolumn{16}{c}{CH$_3$CHO} \\
    \hline
    5$_{\rm 0,5}$--4$_{\rm 0,4}$ E 		&	95.94744	&	13.9	&	-4.5 & 
    3.9(0.2) & 0.4 & 20 &
    $\leq$ 0.6 & $\leq$ 0.06 & 20 & 
    0.8(0.4) & 0.1 & 20 &
    $\leq$0.5 & $\leq$0.05 & 30 \\
    5$_{\rm 0,5}$ - 4$_{\rm 0,4}$ A &	95.96346	&	13.8	&	-4.5 	& 
    3.7(0.3) & 0.4 & 20 &
    1.7(0.4) & 0.1 & 20 & 
    1.3(0.5) & 0.1 & 20 &
     $\leq$0.5 & $\leq$0.05 & 30 \\
    5$_{\rm 2,4}$ - 4$_{\rm 2,3}$ A	&	96.27425	&	22.9	&	-4.6	&
    1.1(0.4) & 0.1 & 30 & 
    0.7(0.3) & 0.07 & 20 & 
    1.0(0.2)& 0.1 & 30 & 
    0.5(0.3) & 0.1 & 20 \\
    5$_{\rm 2,4}$ - 4$_{\rm 2,3}$ E & 	96.42561	&	22.9	&	-4.6 & 
    1.1(0.2) & 0.2 & 30 & 
    $\leq$ 0.7 & $\leq$ 0.07 & 20 &
    0.8(0.3) & 0.06 & 30 & 
    $\leq$0.3 & $\leq$0.03 & 20 \\
    5$_{\rm 2,3}$ - 4$_{\rm 2,2}$ E	&	96.47552	&	23.0	& -4.6	& 
    0.8(0.2) & 0.1 & 20 & 
    $\leq$ 0.7 & $\leq$ 0.07 & 20 &
    $\leq$ 0.8 & $\leq$ 0.08 & 30 &
    $\leq$0.3 & $\leq$0.03 & 20  \\
    5$_{\rm 2,3}$ - 4$_{\rm 2,2}$ A	 &	96.63266	&	23.0	&	-4.6	& 
    1.5(0.3) & 0.1 & 20 &
    $\leq$ 0.7 & $\leq$ 0.07 & 20 &
    $\leq$ 0.8 & $\leq$ 0.08 & 30 &
    $\leq$0.3 & $\leq$0.03 & 20  \\
    5$_{\rm 1,4}$ - 4$_{\rm 1,3}$ E	&	98.86331	&	16.7	&	-4.5	& 
    2.3(0.4) & 0.2 & 20 &
     1.6(0.8) & 0.1 & 20 &
     2.1(0.5) & 0.2 & 30 &
     2.8(0.6) & 0.1 & 20 \\
    5$_{\rm 1,4}$ - 4$_{\rm 1,3}$ A	&	98.90094	&	16.5	&	-4.5	& 
    2.5(0.5) & 0.2 & 20 &
    $\leq$ 0.6 & $\leq$ 0.06 & 20 &
    1.9(0.3) & 0.2 & 30 & 
    2.0 (0.3) & 0.1 & 20 \\
    \hline   
    \multicolumn{16}{c}{{CH$_3$OCH$_3$}} \\
    \hline
     4$_{\rm 1,4}$ - 3$_{\rm 0,3}$ EA $^{(e)}$ &	99.32443	&  10.2 & -5.3 &
     \multirow{4}{*}{$\leq$ 0.2 } &  \multirow{4}{*}{$\leq$ 0.02} &  \multirow{4}{*}{15} &
     \multirow{4}{*}{$\leq$ 0.3 } &  \multirow{4}{*}{$\leq$ 0.03} &  \multirow{4}{*}{15} &
      \multirow{4}{*}{2.1(0.3)} &  \multirow{4}{*}{0.12} &  \multirow{4}{*}{15} &  
     \multirow{4}{*}{1.8(0.5)} &  \multirow{4}{*}{0.09} &  \multirow{4}{*}{15} \\ 
    4$_{\rm 1,4}$ - 3$_{\rm 0,3}$ AE $^{(e)}$ &	99.32443	&  10.2 & -5.3 &
    & & &
    & & &
    & &  \\ 
    4$_{\rm 1,4}$ - 3$_{\rm 0,3}$ EE $^{(e)}$ &	99.32521	&  10.2 & -5.3 &
    & & & 
    & & &
    & & \\ 
    4$_{\rm 1,4}$ - 3$_{\rm 0,3}$ AA $^{(e)}$&	99.32607	&  10.2 & -5.3 &
    & & & 
    & & &
    & &  &
    & & \\
    \hline
   \multicolumn{16}{c}{{NH$_2$CHO$^{(f)}$}} \\
    \hline
    4$_{\rm 1,4}$ - 3$_{\rm 1,3}$  & 81.69354 & 12.8 & -4.5 &
    $\leq$ 0.2 & $\leq$ 0.02 &  7 & 
    $\leq$ 0.1  &  $\leq$ 0.01 &  7 &
    $\leq$ 0.06  &  $\leq$ 0.006 &  7 &
     0.4(0.1)  &  0.04 & 10 \\
    \hline
    \end{tabular}
    \begin{tablenotes}
    \item[a] {Region where formamide and dimethyl ether emits \citep{ceccarelli_seeds_2017};}
    \item[b] {Frequencies and spectroscopic parameters are retrieved from the JPL \citep[Jet Propulsion Laboratory;][]{pickett_submillimeter_1998} molecular database and from the CDMS \citep[Cologne Database for Molecular Spectroscopy;][]{muller_cologne_2005}: for CH$_3$OH by \citet{xu_torsion_2008}, for CH$_3$CHO by \citet{kleiner_microwave_1996}, for CH$_3$CHO$_3$ by \citet{neustock_millimeter_1990}, for NH$_2$CHO by \citet{kirchhoff_microwave_1973}. 
    Upper level energies refer to the ground state of each symmetry.}
    \item[c] {Mean velocity--integrated line flux over the whole velocity emission range from the spectra extracted at the CH$_3$CHO and CH$_3$OH transitions at each outflow peaks (see Table \ref{tab:coordinates}). In case of non--detection we report the 3$\sigma$ limit.
    The lines are centered at $\sim$ 3 km s$^{-1}$, $\sim$ 1 km s$^{-1}$, $\sim$ 11 km s$^{-1}$ for outflow SE, SW and North respectively, in agreement with the expected outflow velocity \citep{santangelo_jet_2015} given the WideX channel resolution ($\sim$ 6 km s$^{-1}$).}
    \item[d] The rms is computed over a 200 km s$^{-1}$ band around each line.
    \item[e] {These lines are blended together at the WideX resolution ($\sim$2 MHz), therefore they are not used for the non-LTE analysis.}
    \item[f] {From Setup 1.}
    \end{tablenotes}
    \end{threeparttable}
    }
\end{table*}

\section{Derivation of the column densities and abundance ratios}\label{sec:CDratios}

\subsection{Methanol and acetaldehyde}
We used the detected lines of methanol and acetaldehyde to estimate their column densities in the four positions of Table \ref{tab:coordinates}.
We used the standard rotational diagram method \citep{goldsmith_population_1999}, which assumes local thermodynamic equilibrium (LTE) and optically thin line emission. Note that we checked a posteriori that the latter assumption is valid. Also, 
because the map shown in Figure \ref{fig:Contour map} shows that the emission is more extended with respect to the observation beam, we did not apply any dilution factor. The used error bar of each data point includes the spectral RMS and the calibration error ($\sim$15\%). 

\begin{figure*}
    \centering
    \subfloat[Outflow SE]{\includegraphics[scale=0.37]{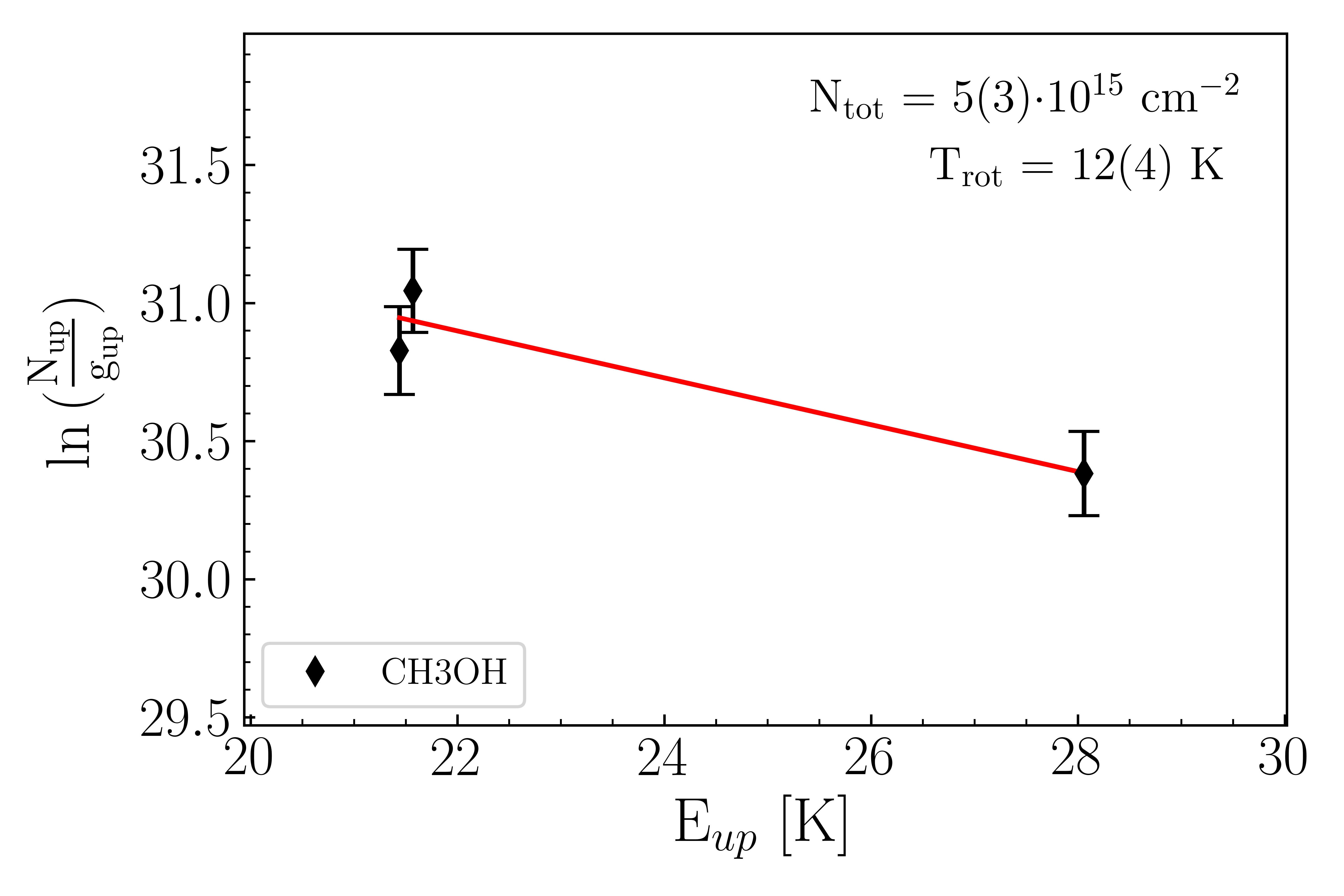}}
    \subfloat[Outflow SW]{\includegraphics[scale=0.37]{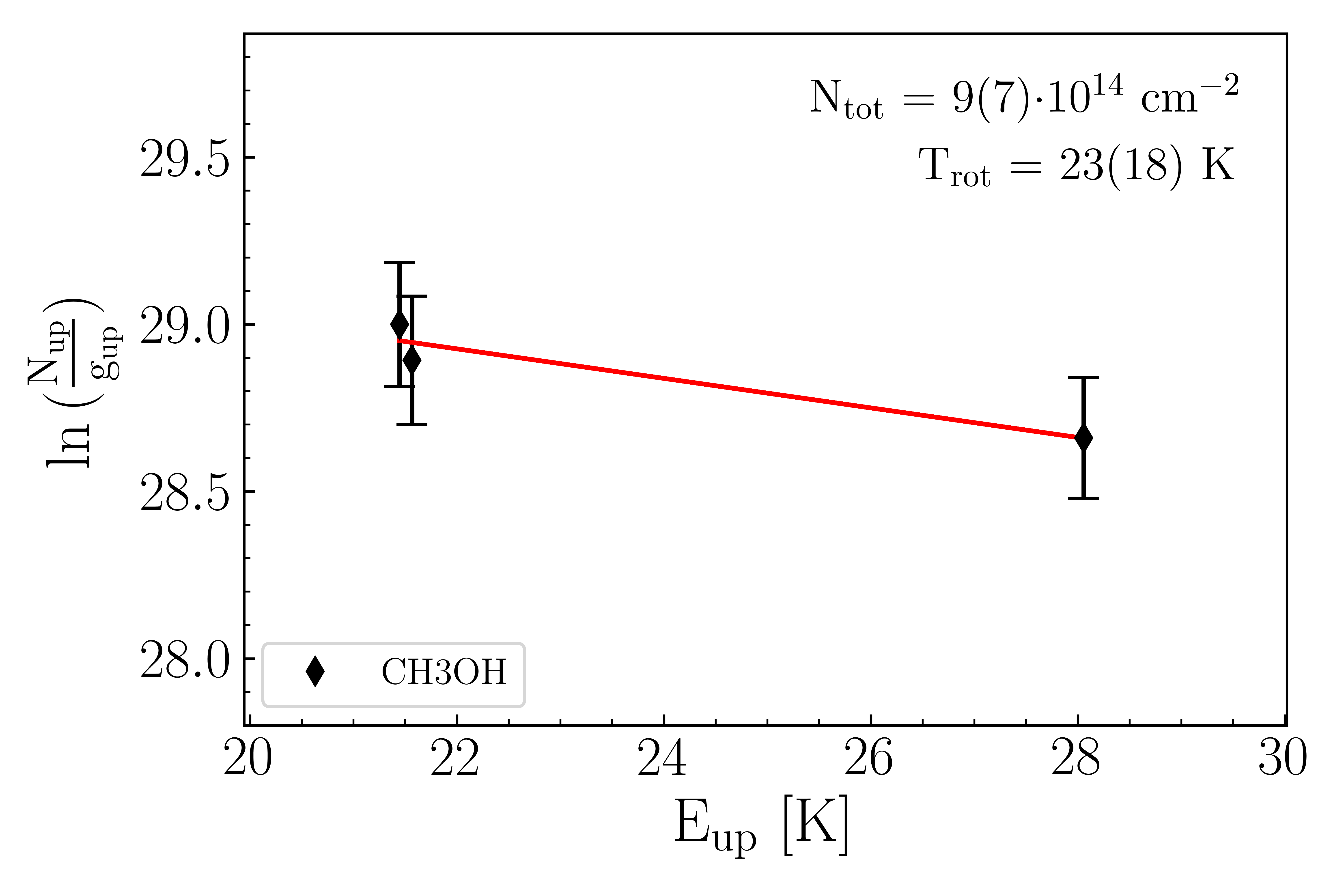}}
    \subfloat[Outflow North]{\includegraphics[scale=0.37]{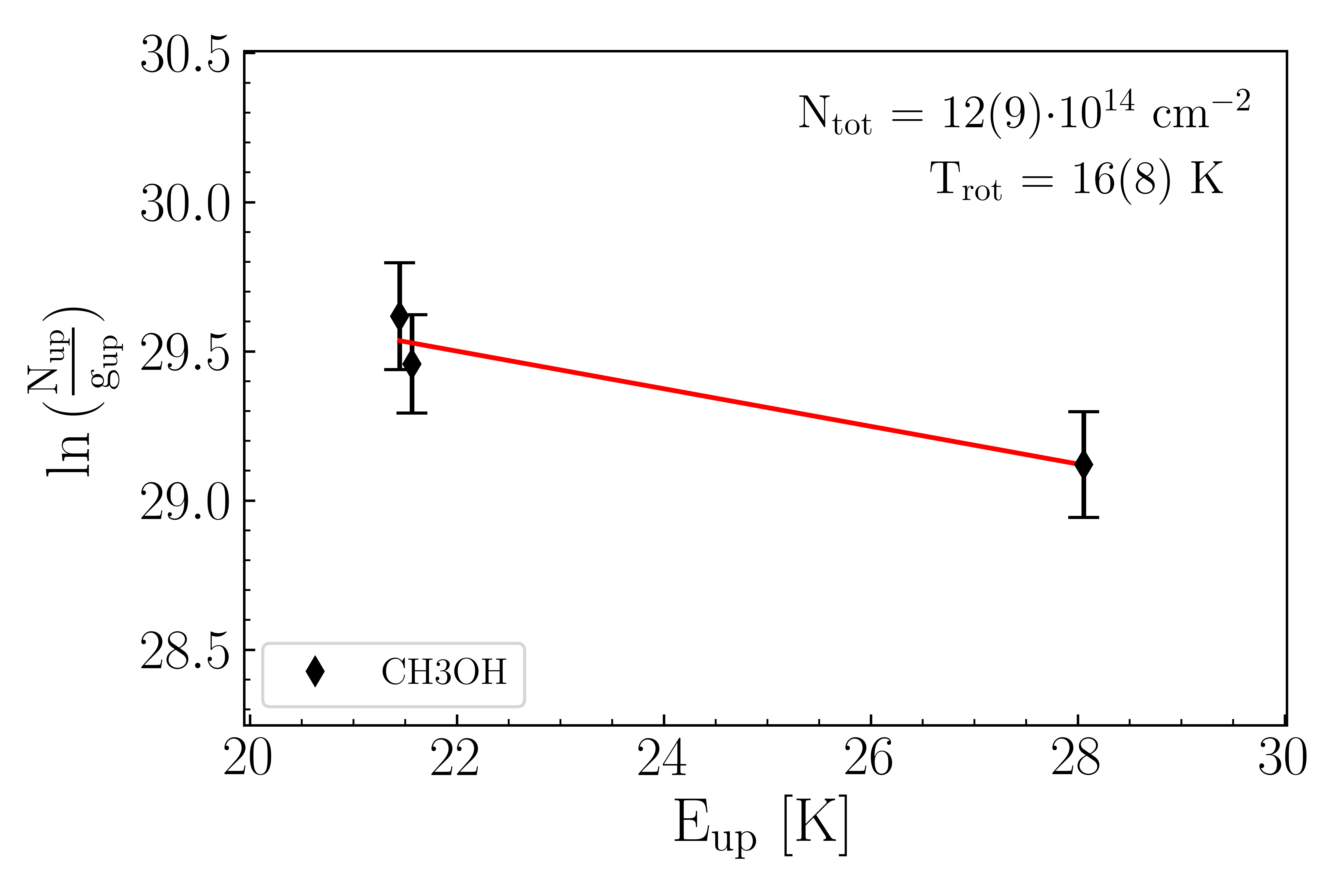}} 
    
    \centering
    \subfloat{\includegraphics[scale=0.37]{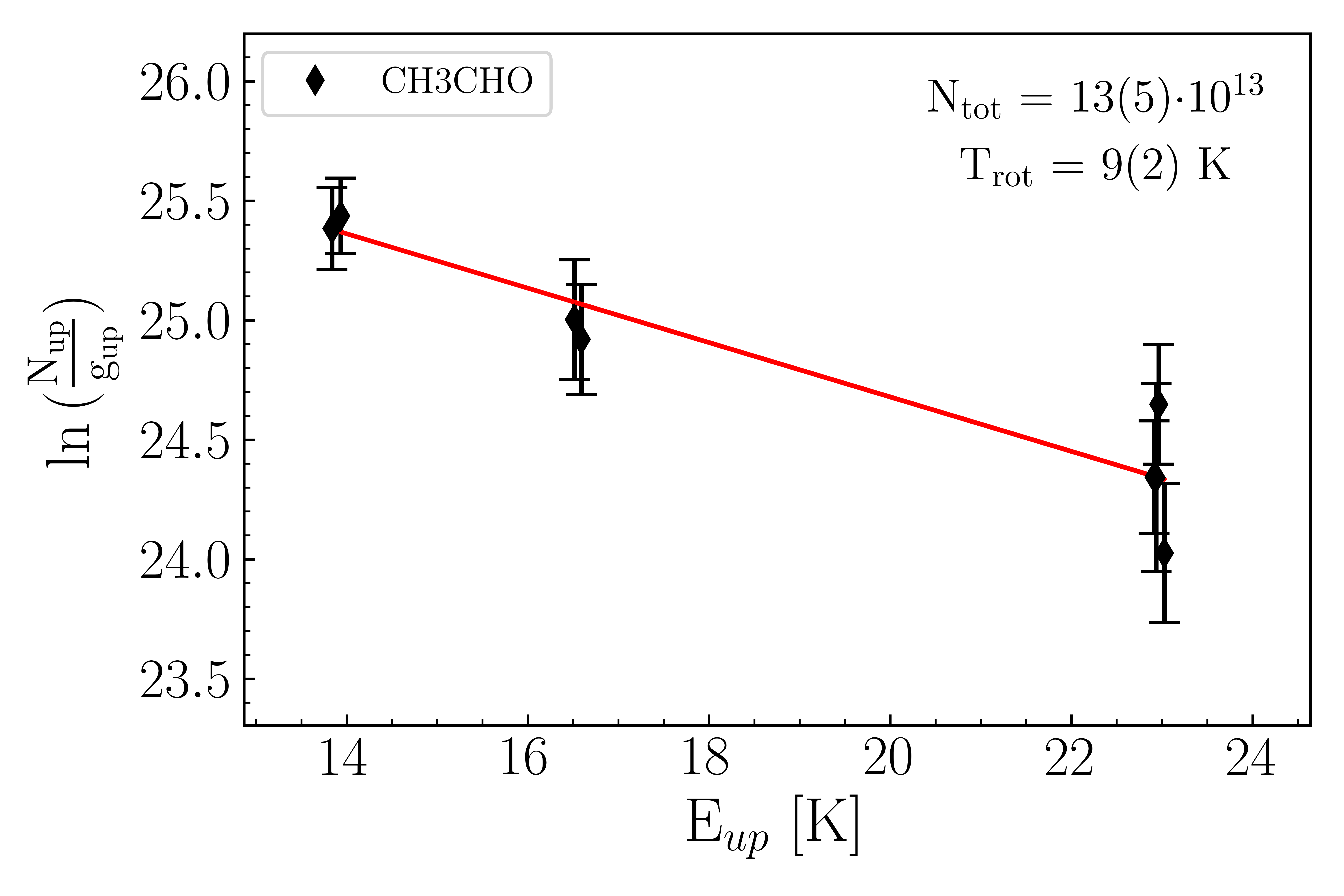}}
    \subfloat{\includegraphics[scale=0.37]{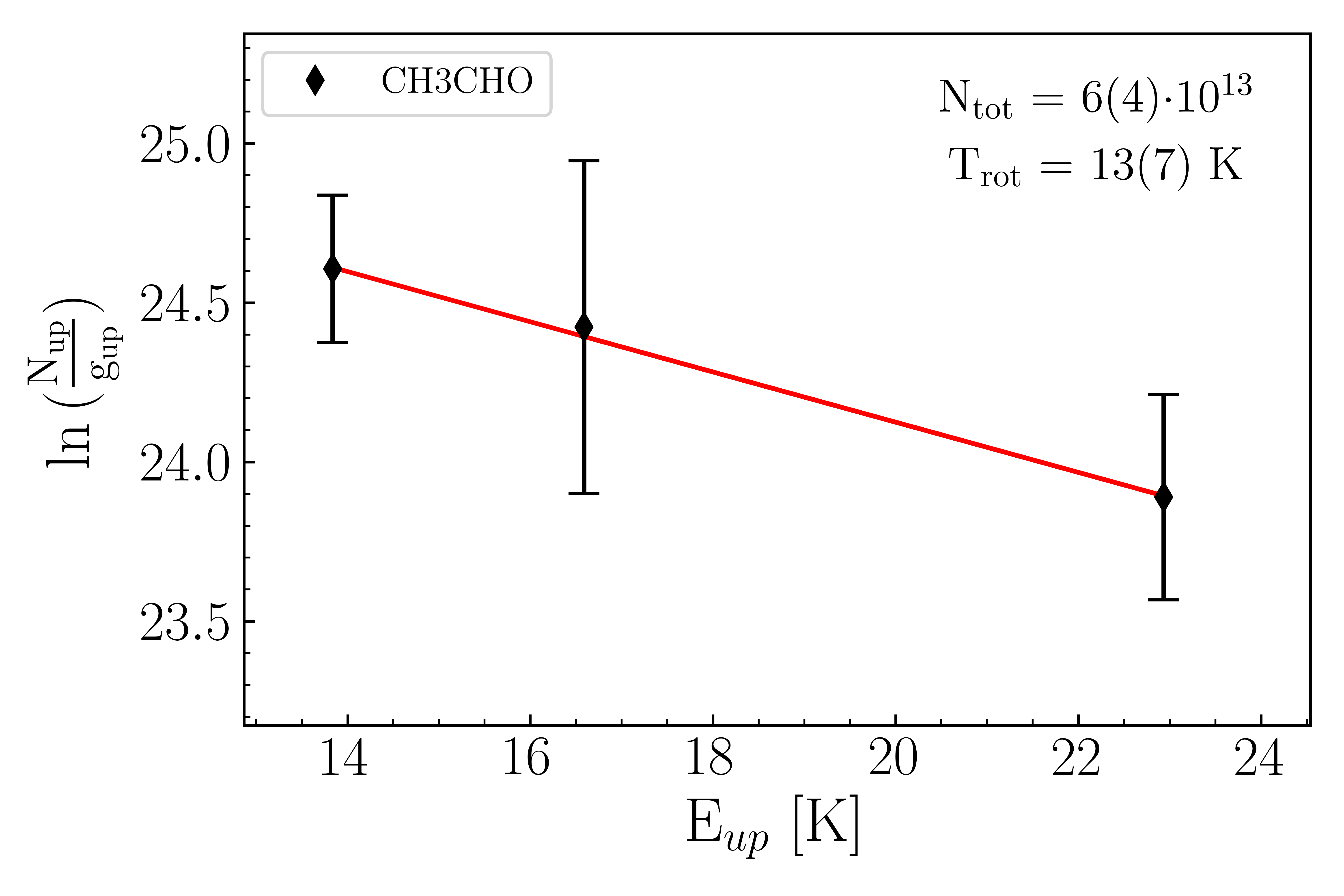}}
    \subfloat{\includegraphics[scale=0.37]{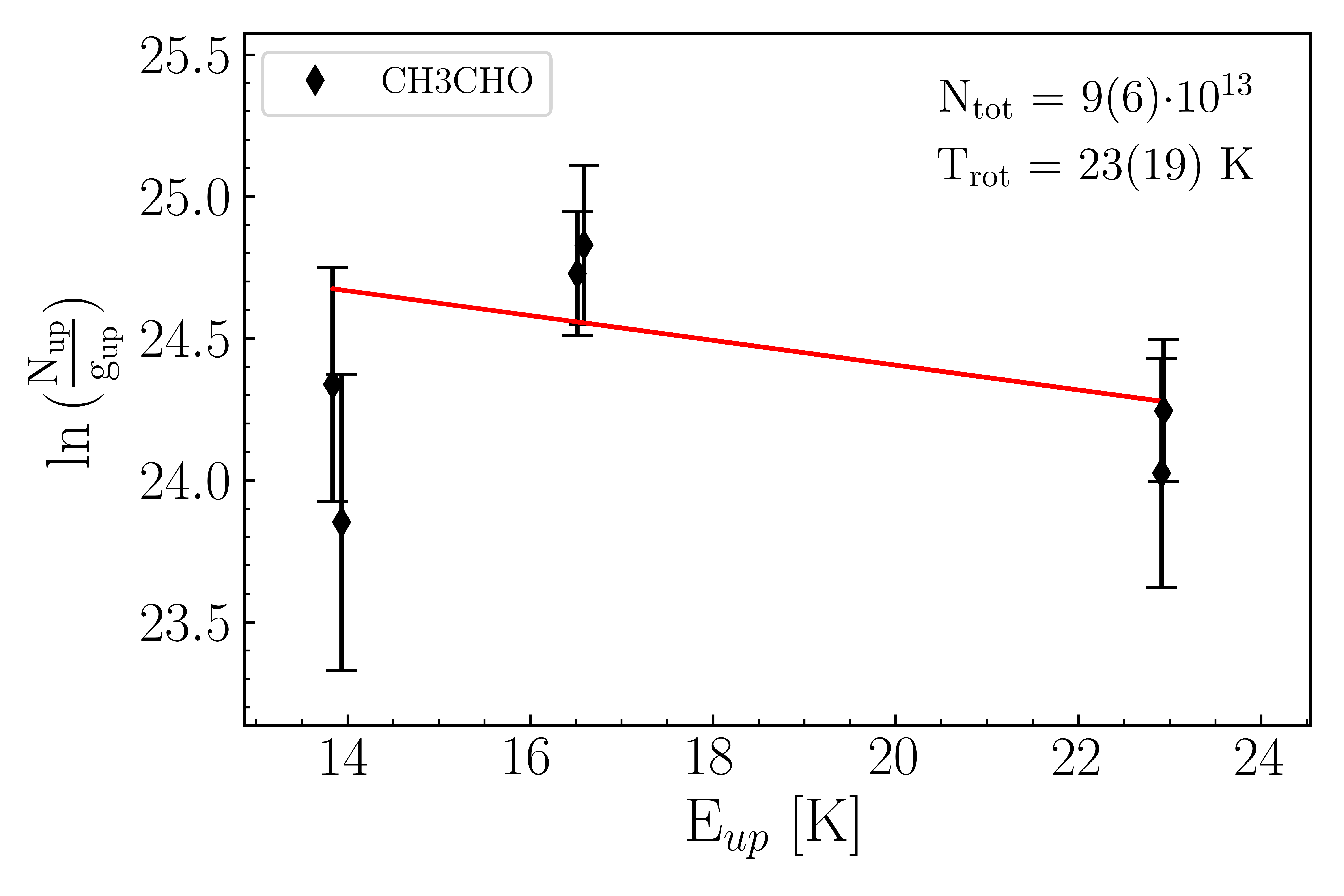}}
    
    \caption{ 
    \textit{Upper panels:} Rotational Diagrams of methanol (CH$_3$OH) in outflow SE, SW and North from left to right;
    \textit{Lower panels:} Rotational Diagrams of acetaldehyde (CH$_3$CHO) in outflow SE, SW and North.
    The parameters N$_{\rm up}$, g$_{\rm up}$, and E$_{\rm up}$ are the column density, the degeneracy, and the energy of the upper level. The error bars on ${\rm \ln(N_{up}/g_{up})}$ are computed by taking the calibration error on the integrated flux (15\%) into account. The red lines represent the best fits. }
    \label{fig:RD plots}
\end{figure*}

Figure \ref{fig:RD plots} shows the rotational diagrams of methanol and acetaldehyde in SE, SW and North positions (Table \ref{tab:coordinates}) and Table \ref{tab:Ntot-Trot-ratio} lists the fitted values.
In the As position, we could not build a rotational diagram  {for either of the two species}, as not enough lines were detected. In this case, we obtained an estimate of their column density, by assuming a rotational temperature ranging from 10 to 30 K (range that includes the temperatures found in the other positions, SE, SW and North). If no line was detected we used the 3$\sigma$ limit.

 {To compute the methanol over acetaldehyde abundance ratios, quoted in Table \ref{tab:Ntot-Trot-ratio}, we adopted the assumptions that the lines emitted by the two species come from the same region and, therefore, possess the same rotational temperature $\rm T_{rot}$. 
Under these two assumptions, the column density ratio $R$ between the two species is obtained by taking the column density $N_x$ of each species at the same rotational temperature $\rm T_{rot}$, namely $R$=$\rm N_1(T_{rot})/N_2(T_{rot})$. The error bar $\delta R$ is then obtained by computing $R$ at the smallest and largest $\rm T_{rot}$ of the two species.
For example, in the Outflow SE the derived $\rm T_{rot}$ is equal to (11$\pm$3) K and (9$\pm$2) K in methanol and acetaldehyde, respectively; therefore, to estimate the error $\delta R$ we computed $R{\rm(T_{rot}})$ at 7 and 14 K. 
Note that the method described above allows to reduce the error bar in the abundance ratio because the calibration uncertainty, which enters in the column density estimate of each species, cancels out when considering the column density of two species derived with the same observation data set.}

For methanol, we obtain a column density N$_{\rm CH_3OH} \simeq 9-50 \times 10^{14}$ cm$^{-2}$ and a rotational temperature T$_{\rm CH_3OH}$ between 11 K and 23 K. 
For acetaldehyde, we obtain N$_{\rm CH_3CHO} \simeq 0.6-1.3 \times 10^{14}$ cm$^{-2}$ and T$_{\rm CH_3CHO}$ between 9 K and 23 K.
Their abundance ratio varies from 10--20 to $\sim$44 in the three lobes; more precisely, the SE lobe is the one with the largest CH$_3$OH/CH$_3$CHO abundance ratio. In As, we only derive a lower limit to the CH$_3$OH/CH$_3$CHO abundance ratio, $\geq$20.
Note that our observations provide different values of the CH$_3$OH/CH$_3$CHO abundance ratio compared to those previously derived by \citet[][$\sim$300]{holdship_observations_2019} towards the south and north lobes of the IRAS 4A outflows via single-dish observations. 
{We attribute this difference to the fact that the single-dish observations of IRAS 4A by \citet{holdship_observations_2019} include emission from a much larger region (including also some from the central protostars) with respect to that probed by the present SOLIS observations. 
Moreover \citet{holdship_observations_2019} do not explicitly derive the emitting size, even though they minimize with respect to this parameter: since the maximization is done independently for each species, their column density ratio has a relatively large intrinsic uncertainty. Therefore the interferometric images allow us to minimize the risk of mixing different gas components (indeed typical of shocked regions).
} 

The abundance ratio analysis confirms what we could see from the emission maps (Figure \ref{fig:Contour map}), namely that the methanol and acetaldehyde emission (and their relative abundance) is quite different in the SE lobe with respect to the SW and Northern lobes. We emphasize that the difference cannot be attributed to excitation effects, having the methanol and acetaldehyde lines with similar upper level energies (E$_{\rm up}$ from 7 K to 28 K), similar Einstein coefficients (A$_{\rm ij}\sim$ 10$^{-5}$ s$^{-1}$) and comparable derived temperature, within the measurement errors. 

To summarize, the southern--west and northern lobes have a relatively low CH$_3$OH/CH$_3$CHO abundance ratio (8--20), whereas the southern--east lobe presents a ratio at least twice larger ($\sim44$). 
Since the SW and North lobes belong to the same outflow emanating from 4A2 and the SE lobe traces the outflow emanating from 4A1, it seems reasonable to attribute the observed difference to an intrinsic difference in the two outflows. We will explore this hypothesis in the next section with the help of an astrochemical model.

\subsection{Dimethyl ether and formamide}
Regarding formamide and dimethyl ether, we detected only one line at most of the latter and four lines (blended at the WideX resolution, $\sim$2 MHz) of the former in any position (see Table \ref{tab:fit_results}. For this reason, we did not carry out the rotational diagram analysis as we did for the other iCOMs. Furthermore, while formamide is marginally resolved at the As position, dimethyl ether is not resolved in any outflow peaks (see Figure \ref{fig:Contour map}). 
In presence of detection, we derived the column densities using the integrated area of the $4_{1,4}-3_{1,3}$ line for formamide and the blended ones for dimethyl ether, assuming fixed rotational temperatures. As for methanol and acetaldehyde, we used 11, 16 and  19 K for SE, SW and North, respectively and 10--30 for As. 
In case of non detection, an upper limit on the column density is derived considering the 3$\sigma$ limit of the spectra and using the above rotational temperatures. 
The derived values are reported in Table \ref{tab:Ntot-Trot-ratio}.

\begin{table*}
    \centering
    \caption{Results of the LTE analysis with the Rotational diagrams, for each outflow peaks, using the observation with the NOEMA WideX backend. In the last row are shown the values of the abundance ratio between methanol and acetaldehyde for each outflow peak.}
    \label{tab:Ntot-Trot-ratio}
    \begin{threeparttable}  
    \begin{tabular}{l|l|c|ccccc}
    \hline
    \hline
    \multirow{2}{*}{Molecule} &\multicolumn{2}{c|}{  } & \multirow{2}{*}{Outflow SE}  & \multirow{2}{*}{Outflow SW} & \multirow{2}{*}{Outflow North} & \multirow{2}{*}{Region As$^{(a)}$} & \multirow{2}{*}{L1157-B1$^{(b)}$} \\
    & \multicolumn{2}{c|}{  } & & & & & \\
    \hline
    \multirow{2}{*}{CH$_3$OH} &  N$_{\rm tot}$ & [10$^{14}$ cm$^{-2}$] & 50(30) & 9(7) & 12(9) &  5-15 & 130(30) \\
     & T$_{\rm rot}$ & [K] & 11(3) & 23(18) & 18(8) & 10-30$^{(c)}$ & 10.0(1.1) \\
  
    \multirow{2}{*}{CH$_3$CHO} &  N$_{\rm tot}$ & [10$^{14}$ cm$^{-2}$] & 1.3(0.5) & 0.6(0.4) & 0.9(0.6) & $\leq$ (0.25-0.5) & 0.7(0.3) \\
     & T$_{\rm rot}$ & [K] & 9(2) & 13(7) & 23(19) & 10-30$^{(c)}$ & 8(1) \\
     
     \multirow{2}{*}{CH$_3$OCH$_3$} &  N$_{\rm tot}$ & [10$^{14}$ cm$^{-2}$] & $ \leq$ 0.5  & $ \leq$ 0.7 & 1.6(0.7) &  1.0(0.4) & 3 \\
     & T$_{\rm rot}$ & [K] & 11 & 16  & 19 & 10-30$^{(c)}$ & 9 \\
      \multirow{2}{*}{NH$_2$CHO} &  N$_{\rm tot}$ & [10$^{14}$ cm$^{-2}$] & $ \leq$ 0.02 & $ \leq$ 0.02 & $ \leq$ 0.02 & 0.03(0.02)& -- \\
     & T$_{\rm rot}$ & [K] & 11  & 16  & 19  & 10-30$^{(c)}$ & -- \\ 
   \hline
     $\rm \dfrac{CH_3OH}{CH_3CHO}^{(d)}$ &  \multicolumn{2}{c|}{  }  & 44(5) & 11(3) & 17(3) & $\ge$ 20  & 190(60)  \\
    \hline
    \end{tabular}
     \begin{tablenotes}
      \small
      \item[a] Region where formamide and dimethyl ether emits \citep{ceccarelli_seeds_2017};
      \item[b] From \citet{codella_seeds_2019} based on interferometric observations;
      \item[c] Fixed rotational temperature used to derive a range of possible N$_{\rm tot}$;
      \item[d]  {Abundance Ratio computed dividing the best fit column densities derived assuming that the two species are tracing the same gas with the same properties (see text)}.
    \end{tablenotes}
    \end{threeparttable}  
\end{table*}

\section{Astrochemical modelling}\label{sec:modelling}

We ran our {model GRAINOBLE+ in order to reproduce} the observations and understand what could be the cause of the observed difference in the CH$_3$OH/CH$_3$CHO abundance ratio in the different lobes of the IRAS 4A outflows.

There is solid (observational, theoretical and from laboratory experiments) evidence that methanol is synthesized on the grain surfaces via the hydrogenation of iced CO by successive addition of H atoms \citep{boogert_observations_2015,tielens_model_1982,watanabe_efficient_2002}.
On the contrary, the acetaldehyde formation route is less clear, and the two paths, formation in the gas phase or on the grain surfaces, are still debated. 
Specifically, grain surface models predict that CH$_3$CHO could be formed through the combination of the two radicals CH$_3$ and HCO (previously formed by photodissociation of methanol and formaldehyde, respectively) on the surface of the grains \citep{garrod_formation_2006}. However, recent quantum chemistry computation by \citet[][]{enrique-romero_impossible?_2016,enrique-romero_reactivity_2019} show that alternative channels leading back to the two simple species CH$_4$ and CO are competitive. 
Conversely, gas phase models claims that acetaldehyde formation could occur by the oxidation of hydrocarbons \citep[formed previously on the grain mantles as hydrogenation of carbon chains; ][]{charnley_molecular_1992,charnley_acetaldehyde_2004}. In particular, the injection from grain mantles of ethane (C$_2$H$_6$) is expected to drive CH$_3$CH$_2$ that will then react in the gas phase with atomic oxygen, giving CH$_3$CHO \citep[][]{charnley_acetaldehyde_2004}. The crucial reaction is, therefore:
\begin{equation}\label{eq:ethyl+O}
\rm CH_3CH_2 +O \rightarrow CH_3CHO + H 
\end{equation}

Following these two possibilities, we ran a grid of astrochemical models in order to reproduce our observations and to understand the possible cause of the difference in the observed CH$_3$OH/CH$_3$CHO values in the two IRAS 4A outflows.

\subsection{Model description}\label{secsub:model_descript}
GRAINOBLE+ is a gas--grain model simulating the chemical evolution of gas and ices. It is the upgraded version of GRAINOBLE, initially developed by \citet{taquet_multilayer_2012}; in particular it is re-coded and improved in terms of computational efficiency and treatment of processes. The GRAINOBLE+ version\footnote{A detailed description of the GRAINOBLE+ model will be reported in a forthcoming dedicated article.} allows an easy incorporation of many complicated processes that occur in the gas and on the grain surfaces. 
The code can carry out an easy implementation of evolution of physical conditions of a cloud with a given time dependent physical profile. Additionally, the new code allows a distribution of size for the grains, multilayer formation of the grain ice mantle, growth and depletion of the ice, and desorption. 

In this work, we used a chemical network of 522 species and 7785 reactions based on the KIDA database\footnote{\url{http://kida.obs.u-bordeaux1.fr}} which has been updated from various recent works \citep[e.g.][]{loison_interstellar_2014,balucani_formation_2015,skouteris_new_2017,skouteris_genealogical_2018}.

In order to simulate the passage of a shock in IRAS 4A, we employed the gas--phase mode of GRAINOBLE+. The simulation follows two steps: (1) a cold molecular gas phase at 10 K and  $2 \times 10^4 \text{cm}^{-3}$ H--nuclei density ($\text{n}_\text{H}$); (2) a post--shock gas phase where density and temperature suddenly jump to $2 \times 10^6 \text{cm}^{-3}$ and 70 K. 

In other words, the second phase inherits the evolved chemical composition of the cloud from the cold phase\footnote{Note that we do not compute the grain mantle composition as this is treated as a parameter in the following analysis.}. Additionally, the gas is infused by species that were formerly synthesized in ice mantles due to the grain sputtering caused by the shock passage.

It is possible that, before reaching the temperature of 70 K, the shocked gas passes through a short initial phase with high temperature.
However, this phase unlikely affects the results reported in Fig. \ref{fig:model_timeevol} because there are not known gas--phase reactions with activation barrier forming acetaldehyde. 
 {This is confirmed by models taking into account the temperature evolution in the shocked gas \citep[e.g.][]{nesterenok_chemical_2018, burkhardt_modeling_2019}}.

We do not have precise estimates of the density and temperature in the positions where we derived the CH$_3$OH/CH$_3$CHO abundance ratio. 
However, the values derived  {at positions close to the ones selected here}, via a non-LTE analysis of observed SO$_2$ lines, \citep{taquet_seeds_2019} are similar to those 
adopted in our modelling.
We assumed that the cosmic ray  {ionization} rate is the same than towards L1157-B1 \citep[$\rm \zeta=3\times 10^{-16} \ s^{-1}$,][]{podio_molecular_2014} and that the abundance of the injected species are also similar to the ones adopted for L1157--B1 \citep[see Table B1 of][]{codella_seeds_2017}, listed in Table \ref{tab:injected_abundances}.

We ran series of models to compare the observed CH$_3$OH/CH$_3$CHO abundance ratio with the predicted one and to understand what the ratio depends on.
We start exploring the gas phase formation route of acetaldehyde assuming that its formation is dominated by reaction \ref{eq:ethyl+O} in the passage of shock. 
Oxygen is much more abundant than the injected ethyl radical in the post--shock phase; therefore, CH$_3$CH$_2$ is the bottleneck of the rate of the reaction \ref{eq:ethyl+O}.

First, we ran a grid of 169 models with different injected abundances in the beginning of the shocked phase for ethyl radical (CH$_3$CH$_2$), and methanol from $[4 \times 10^{-9}, 4 \times 10^{-7} ]$ and $[4 \times 10^{-8}, 4 \times 10^{-6}]$ ranges respectively. The methanol abundance range is chosen in order to include the observed values in the hot corinos and in the protostellar shocks (the L1157-B1 outflow in particular); the ethyl radical one is chosen in order to match our observed CH$_3$OH/CH$_3$CHO abundance ratio. The results are shown in Figure \ref{fig:model_contour} and described in the next section.

Second, we studied the influence of the density and cosmic ray ionization rate $\zeta$,  on the chemical evolution and how the CH$_3$OH and CH$_3$CHO abundances and their relative ratio depend on the time, after the shock passage.
Note in fact that protostellar shocks could be local accelerators of cosmic ray protons \citep{padovani_protostars:_2016} and therefore, in this work $\zeta$ is an unknown parameter.
We ran then two additional models. In the first one,  {we lower down the} chosen n$_\text{H}$ density value up to $2 \times 10^5$ cm$^{-3}$; in the second one,  {we lower down the chosen} cosmic ray ionization rate value up to $\zeta=3 \times 10^{-17}$ s$^{-1}$ for the post--shock phase. The injected abundances of methanol is $3 \times 10^{-6}$ and for ethyl radical it is $3 \times 10^{-7}$ in both cases. These values are chosen such that the models reproduce the observed CH$_3$OH/CH$_3$CHO abundance ratios for the SW and North lobes (8-20) at 1000 yr (orange band of Figure \ref{fig:model_contour}) and for the SE one as well.
The results are shown in Figure \ref{fig:model_timeevol} and described in the next section.

Finally, we ran a model with the assumption that acetaldehyde is synthesized on the ice mantles and injected directly into the gas phase at the passage of the shock. 

\begin{table}
    \centering
    \caption{The injected abundances (with respect to H-nuclei) into the gas phase at the second step of the model. These values are based on previous observations towards L1157-B1 \citep{codella_seeds_2017}.}
    \label{tab:injected_abundances}
    \begin{tabular}{l|c}
    \hline
    \hline
     molecules  & injected abundances $\rm \left(/H \right)$ \\ 
     \hline
     CO$_2$ & $3 \times 10^{-5}$  \\
     H$_2$O & $2 \times 10^{-4}$  \\
     OCS & $2 \times 10^{-6}$  \\
     H$_2$CO & $1 \times 10^{-6}$  \\
     NH$_3$ & $2 \times 10^{-5}$  \\
     \hline
    \end{tabular}
\end{table}

\subsection{Model results}

\begin{figure}
    \centering
    \subfloat{\includegraphics[scale=0.6]{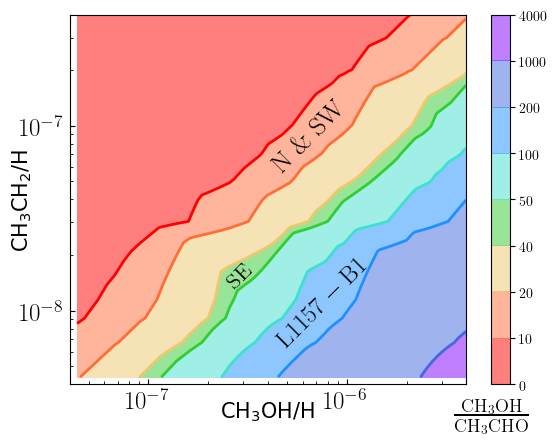}}    
    \caption{ 
    Contour map of the CH$_3$OH/CH$_3$CHO abundance ratio at 1000 year after the start of the shock passage. The x and y axis are the injected abundances of methanol $[4 \times 10^{-8}, 4 \times 10^{-6}]$ and the parent molecule of acetaldehyde, ethyl radical CH$_3$CH$_2$ $[4 \times 10^{-9}, 4 \times 10^{-7}]$, respectively. The measured methanol to acetaldehyde abundance ratio of North and SW outflows $\left(8-20\right)$ fall in the orange band and the SE one $\left(38-50\right)$ in the green band; L1157-B1 value $\left(130-250\right)$ is covered by the blue band \citep{codella_seeds_2019}.
    }
    \label{fig:model_contour}
\end{figure}

\begin{figure}    
    \centering
    \subfloat{\includegraphics[scale=0.6]{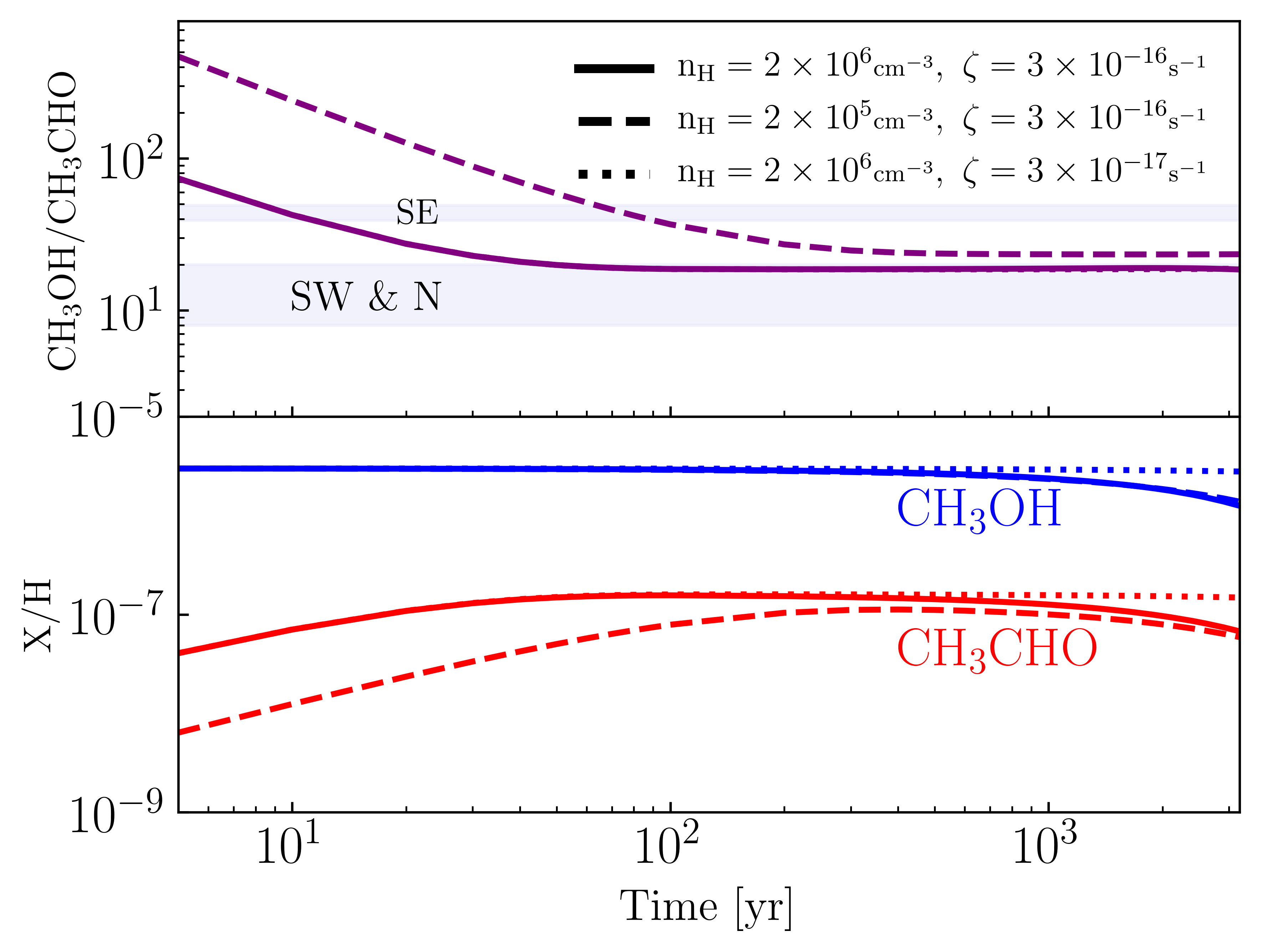}}
    \caption{ 
    Time evolution of abundances of methanol in blue and acetaldehyde in red (bottom) and their ratios in purple (top) for the same injected methanol abundance, $3 \times 10^{-6}$, and ethyl radical one, $3 \times 10^{-7}$.  The different line styles correspond to models run with different conditions, as reported in the upper panel legend. 
     {Please note that the CH$_3$OH/CH$_3$CHO abundance ratio is constant if both species are directly injected from the grain mantles (see text)}.
    }
    \label{fig:model_timeevol}
\end{figure}

Figure \ref{fig:model_contour} shows the contour map of the CH$_3$OH/CH$_3$CHO abundance ratio at 1000 years after the shock passage, as a function of the injected methanol and ethyl radical, assuming that acetaldehyde is entirely synthesized in the gas phase.
Note that the chosen age (1000 yr) is in the order of magnitude of the kinematical age of L1157-B1 \citep{podio_first_2016, codella_seeds_2017} and, likely, IRAS 4A outflows.
First, the figure shows that there is a linear dependence of the CH$_3$OH/CH$_3$CHO abundance ratio both on the injected methanol and ethyl radical abundance, in the range explored in our simulations. 
Second, reasonable values of methanol and ethyl radical abundance can reproduce the observed values in the four IRAS 4A outflow positions that we studied (Table \ref{tab:coordinates}).

In Figure \ref{fig:model_timeevol}, we show the evolution of methanol and acetaldehyde abundance as a function of the time after the shock passage, for different densities $\rm n_H$ and cosmic ray  {ionization} rates $\zeta$.
Given that methanol is a grain surface product, even by varying $\rm n_H$ or $\zeta$, the injected abundance remains constant in the early stages of the shock up to $\sim 2 \times 10^3$ years.
The time evolution of acetaldehyde in the early stages is different from the one of methanol; while the latter remains constant, acetaldehyde abundance increases soon after the shock passage. This is expected, since we assumed that acetaldehyde is a gas phase product and its evolution is dominated by the formation through ethyl radical reaction (\ref{eq:ethyl+O}).
Furthermore, the time evolution of acetaldehyde in the post--shock phase is influenced by variation of $\rm n_H$ or $\zeta$. 
Lower density slows down the formation rate of acetaldehyde in the early stages; this is due the fact that in lower density gas the ion abundance is higher. Therefore, the contribution of destruction rates is higher in the evolution. 
Similar to methanol, a lower cosmic ray  {ionization} rate does not affect the early stages.

 {Finally, we ran a model in which acetaldehyde is synthesized on the icy dust surfaces and injected directly into the gas phase. Therefore, in the model, the species injected into the gas phase right after the shock, is not anymore the ethyl radical but the acetaldehyde, whose abundance is chosen to be equal to the observed one.
Therefore, the gas--phase production of acetaldehyde is, in this case, negligible. } 
As for methanol, the abundance of acetaldehyde, now injected from grains, remains constant up to $\sim2 \times $10$^3$ yr when the destruction by ions (H$_3$O$^+$ in this specific case) becomes dominant. 
Therefore, the difference in the abundance of acetaldehyde between the gas-- and grain-- synthesized is only within the first 200 years of the shocked phase, when acetaldehyde takes time to be formed by the reactions between ethyl radical with atomic oxygen.

\section{Discussion}\label{sec:discussion}

\subsection{The two IRAS 4A outflows}\label{sec:4A-Outflows}

Our new SOLIS observations show that the two outflows emanating from 4A1 and 4A2, previously identified by \citet{choi_variability_2005} and \citet{santangelo_jet_2015} via simple molecules (SiO, CO and SO), are enriched with methanol all across the three (visible) lobes. Conversely, acetaldehyde is spread only over the southern lobes and is concentrated in a compact spot in the northern one.
Therefore, a first conclusion of this work is that, in the protostellar post--shock regions there is the release in gas phase of methanol, previously formed on the grain surfaces, and the production of acetaldehyde. The statistic is very poor for the moment, with the two IRAS 4A outflows and the L1157--B1 one in which acetaldehyde is detected with interferometric observations \citep[][]{codella_astrochemistry_2015, codella_seeds_2017, codella_seeds_2019}.

Regarding the IRAS 4A outflows, the SE lobe is richer in both methanol and acetaldehyde, namely the outflow emanating from 4A1. 
We notice that this is an opposite behavior with respect to the SiO emission, which is instead brighter along the 4A2 outflow \citep{choi_variability_2005}. This anti-correlation with SiO is also seen in other (simple) molecules, such as NH$_3$, H$_2$CO and HCN by \citet{choi_radio_2011}. As \citeauthor{choi_radio_2011} suggested, it could be due to a different strength of the shock (the sputtering of Si could require different shock velocities with respect to the other molecules) or because the SiO traces different physical conditions with respect to the other molecules. For example, SiO could trace the jet while the other molecules could be originated in the gas entrained by the jet \citep[e.g.][]{bachiller_molecular_1998,ospina-zamudio_first_2018,ospina-zamudio_molecules_2019}. 
Linked with this, the different spatial distribution between SiO and other species, could be due to time-evolution effects, namely different ages of the shocks, as previously observed in other outflows \citep[e.g.][]{castets_multiple_2001}. We will discuss more this point in the next section.

Finally, it is not clear what is the origin of the iCOMs emission in the As position. It does not seem to be clearly associated with any of the two southern lobes but rather with a point where they intersect. That would imply that additional shocks can occur at the interface of the swept-up cavities opened up by the jets. Higher spatial resolution observations are needed to confirm or reject this hypothesis.

\subsection{The CH$_3$OH/CH$_3$CHO abundance in IRAS 4A1 and 4A2 outflows}\label{sec:abu-metha-ace}

In Section \ref{sec:CDratios}, we measured the methanol over acetaldehyde abundance ratio towards the three positions of the two IRAS 4A outflows, in the North, SW and SE lobes, respectively (Table \ref{tab:Ntot-Trot-ratio}). While the North and SW lobes have a similar values, between 8 and 20 (considering the error bars), the SE lobe has larger CH$_3$OH/CH$_3$CHO, 38--50. 
In other words, the methanol over acetaldehyde abundance ratio is about twice larger in the outflow emanating from 4A1 with respect to the one from 4A2. In this section, we try to understand the origin of this difference, having in mind that, while methanol is a past grain-surface product, acetaldehyde can either be itself a past grain-surface or a gas-phase product.

One easy possible explanation, then, of the CH$_3$OH/CH$_3$CHO difference is that the grain mantle composition is different in the two outflows. However, this seems unlikely, because, on one hand, no gradient in the ratio is seen between the North and SW lobes of the 4A2 outflow; moreover, the SW and SE lobes are very close in space, closer than the two 4A2 outflow positions where we estimated the CH$_3$OH/CH$_3$CHO abundance ratio. Therefore, although we cannot totally exclude it, it seems to us that the different grain composition is an improbable explanation.

If acetaldehyde is synthesized by the gas-phase reaction (\ref{eq:ethyl+O}), there are more possibilities other than a different grain mantle composition. As shown by the modelling of section \ref{sec:modelling}, a smaller density or a younger age of the 4A1 outflow with respect to the 4A2 one would explain the observed CH$_3$OH/CH$_3$CHO difference. Specifically, if the two outflows are very young and 4A1 is younger than about 200 yr, then this would explain why its ratio is larger than the 4A2 one (see Figure \ref{fig:model_timeevol}).

Unfortunately, our observations did not have enough methanol lines to allow a meaningful non-LTE analysis to derive the volume density, so we do not know if the density in the 4A1 outflow is lower than in the 4A2 outflow. With a non-LTE analysis on SO$_2$, \citet{taquet_seeds_2019} suggest that there is no significant difference in density between the outflow driven by 4A1 and the one from 4A2. 
On the other hand, assuming a typical shock velocity of 100 km/s, we estimate a kinematical age of $\sim$200 yr for the 4A1 outflow which has a very short extent; this seems to support the younger age of 4A1 hypothesis. 
Furthermore, \citet{santangelo_jet_2015}, using high spatial resolution observations of CO, SiO and SO, showed that the 4A1 jet is faster than the 4A2 one; this, combined with the smaller spatial extension again support the hypothesis that 4A1 outflow is younger than the 4A2 one.
In favor of a different age of the two outflows there is also the observed chemical differentiation between the two driving sources, 4A1 and 4A2: the former is bright in the continuum but lacks iCOMs line emission, exactly the opposite of 4A2 \citep[e.g.][]{lopez-sepulcre_complex_2017}. One of the possible explanation for this situation is the smaller hot corino size, which could also imply a younger age of 4A1, and this agrees with the younger age of its outflow too.

In summary,the different CH$_3$OH/CH$_3$CHO abundance ratio measured in the 4A1 and 4A2 outflows is unlikely caused by a different grain mantle composition of the two outflows, because the more extended 4A2 outflow shows no significant variation of this ratio over a scale of about 6000 au. On the contrary, the observed CH$_3$OH/CH$_3$CHO abundance ratio is consistent with the scenario in which (i) the 4A1 outflow is younger  {(and, consequently, faster)} than the 4A2 one and (ii) in both outflows acetaldehyde is synthesized in the gas phase. The major reaction is that between atomic oxygen and ethyl radical. The gas-phase synthesis hypothesis also agrees with theoretical quantum chemistry studies \citep{enrique-romero_impossible?_2016, enrique-romero_reactivity_2019}.
Thus, although terrestrial laboratory experiments show that acetaldehyde can be formed on the surfaces of dust grains \citep[e.g][]{bennett_combined_2005, bennett_laboratory_2005, oberg_formation_2009}, our results provide evidence that the gas phase formation route cannot be neglected and actually appears to be the dominant one in the IRAS 4A outflows.
 {We emphasize that these conclusions are robust, as they very little depend on the details of the modeling being based on known reactions in the gas-phase. }

\subsection{Comparison with other Solar-type objects}\label{sec:comparison}

\begin{figure}
\centering
\includegraphics[scale=0.65]{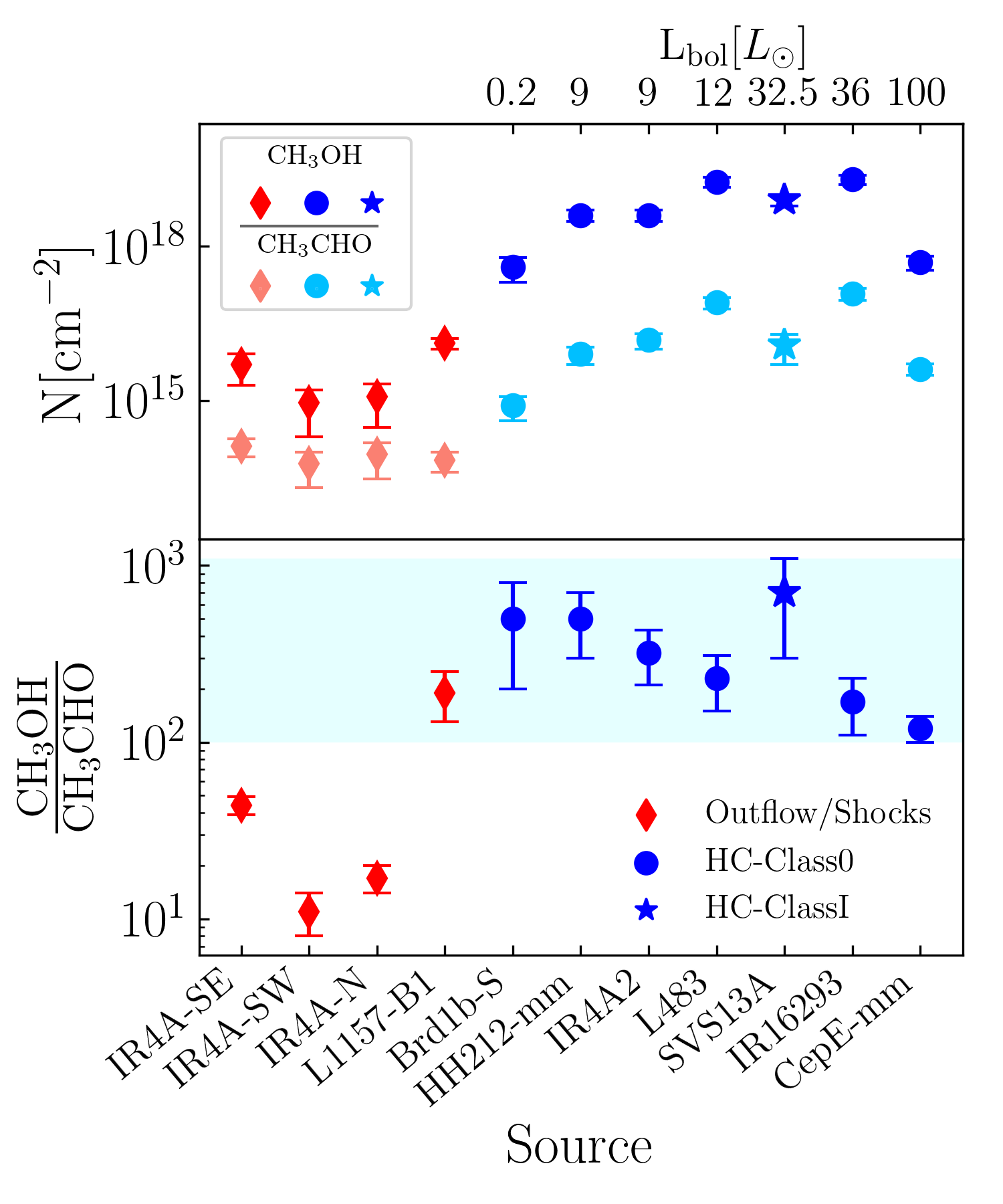}
\caption{Abundance ratios between CH$_3$OH and CH$_3$CHO (\textit{Bottom panel}), CH$_3$OH and CH$_3$CHO column density (\textit{Upper panel}) compared to different sources {for which the emitting size has been estimated via interferometric observations}: the values in the outflows of IRAS 4A (this paper), in the outflow of L1157-B1 \citep{codella_seeds_2019}, the Class I source SVS 13A \citep{bianchi_census_2019}, and the Class 0 sources, in decreasing order of bolometric luminosity, Barnard1b-S \citep{marcelino_alma_2018}, HH212-mm \citep[][]{bianchi_deuterated_2017, codella_seeds_2019}, IRAS 4A2 \citep{taquet_constraining_2015, lopez-sepulcre_complex_2017}, L483 \citep{jacobsen_organic_2018}, IRAS 16293-2422B \citep{jorgensen_alma_2016}, CepE-mm \citep{ospina-zamudio_first_2018}. The outflow values are shown with red (pink for CH$_3$CHO) markers, the hot corinos in blue (cyan for CH$_3$CHO), using diamonds for the outflows, dots for Class 0 and star for the Class I source. The cyan band represents the area in which the the hot corinos values fall.  }
\label{fig:ratio_MetAce_vs_Object}
\end{figure}

Figure \ref{fig:ratio_MetAce_vs_Object} shows the methanol over acetaldehyde abundance ratio in the IRAS 4A outflows, compared with the values measured in other low/intermediate star forming regions for which the emitting size as been estimated {via interferometric observations}: the L1157-B1 molecular shock and seven hot corinos.

First, the IRAS 4A outflows have a definitively lower CH$_3$OH/CH$_3$CHO with respect to the  L1157--B1 value (130-250), reliably measured by  \citet{codella_seeds_2019}. Following the discussion of the previous section, this could be due to a difference in the composition of the grain mantles, to a lower density or to a younger age of the 4A outflows with respect to the L1157-B1 one. At present, the information in our hands is not enough to be able to support or rule out any of these three possibilities; a more accurate analysis of the IRAS 4A outflows is needed. Obviously, having more information of one key actor, the abundance of ethyl radical in these outflows, could shed more light on why the two outflow systems are different.

Finally, the same difference observed between the IRAS 4A and L1157 outflows, if not larger, is observed with respect to the values measured in Class 0 and I hot corinos \citep[][]{marcelino_alma_2018,bianchi_deuterated_2017,taquet_constraining_2015, lopez-sepulcre_complex_2017,jacobsen_organic_2018,jorgensen_alma_2016,bianchi_census_2019, codella_seeds_2019,ospina-zamudio_first_2018}. 

In Figure \ref{fig:ratio_MetAce_vs_Object}, we ordered the hot corinos according their (increasing) bolometric luminosity L$_{\rm bol}$, keeping in mind that the measure of L$_{\rm bol}$ is relatively uncertain.
It is possible to notice an interesting trend: in the Class 0 hot corinos,the CH$_3$OH/CH$_3$CHO abundance ratio decreases with increasing L$_{\rm bol}$. 
Our modeling, (\S \ref{sec:modelling}), is not directly applicable to the hot corino case for two main reasons: 1) the hot corinos density (n$_{\rm H}\sim$10$^7$cm$^{-3}$) is larger than those used in our model (Figure \ref{fig:model_timeevol}, and 2) there is continuous injection of the infalling material towards the center. Having said that, it seems improbable that the behavior shown in Figure\ref{fig:ratio_MetAce_vs_Object} has something to do with a time effect. In the hot corino conditions (higher density and low cosmic  {ionization} rate) the synthesis in the gas phase is fast during the first decades while the destruction by molecular ions is slow in the latest stages($\geq 10^5$ yr), leading to a CH$_3$OH/CH$_3$CHO abundance ratio almost constant. 

On the contrary, a possible interpretation is that larger luminosity correspond to larger hot corino sizes \citep[namely larger regions with a dust temperature $\geq$100 K;][]{ceccarelli_extreme_2007} and, assuming a central peaked density distribution, regions with lower densities. Since methanol is produced during the prestellar phase by hydrogenation of frozen CO \citep{taquet_formaldehyde_2012,vasyunin_formation_2017}, a larger density will bring a larger methanol abundance; acetaldehyde, on the other hand, could be either a past grain-surface or a present-day gas-phase product (see \S \ref{sec:modelling}).
Therefore, the decreasing CH$_3$OH/CH$_3$CHO abundance ratio could indicate that, while methanol abundance decreases with density, acetaldehyde or its gas-phase precursors do not. We emphasize that these conclusions have to be taken with caution as the errors on the hot corinos bolometric luminosity are relatively large, but they are worth a deeper study.

\section{Conclusions}\label{sec:conclusions}
In this work we reported new observations using the IRAM/NOEMA interferometer in the context of the SOLIS large program, and the detection of several iCOMs in the two outflows emanating from IRAS 4A1 and 4A2, respectively: methanol, acetaldehyde, dimethyl ether and formamide. This is the second ever outflow system, after the Solar-type protostellar L1157 outflow, where multiple iCOMs have been detected using interferometers. 
Our main conclusions are the following:
\begin{itemize}
    \item As in the case of L1157-B1 \citep{codella_seeds_2017}, iCOMs are not homogeneously distributed across the IRAS 4A outflows: methanol is more widespread over the two outflows, while acetaldehyde emission is only bright in the southern lobes, and dimethyl ether and formamide are concentrated in a spot at the (apparent) interface between the south lobes of the 4A1 and 4A2 outflows. We, therefore, forewarn that derivation of iCOMs abundance ratios from single-dish observations could be hazardous. 
    \item The measured methanol over acetaldehyde abundance ratio is twice larger in the 4A1 outflow with respect to that in the 4A2 one; the comparison between these results with model predictions suggests that: 1) the 4A1 outflow is younger than the 4A2 one and 2) acetaldehyde is synthesized in the gas phase by the reaction of atomic oxygen with ethyl radical. 
    Alternatively, the grain mantle distribution should vary widely on small scale, which seems unlikely since the larger scale 4A2 outflow shows a similar CH$_3$OH/CH$_3$CHO in two distant points of the southern and northern lobes.
    \item Considering the CH$_3$OH/CH$_3$CHO abundance ratio, the two IRAS 4A outflows show a sharp difference with respect to the L1157-B1 one. This may indicate that either the grain mantles or the gas volume densities are very different in the two regions. Additional observations are necessary to better constrain the reason of the observed difference.
    \item The methanol over acetaldehyde abundance ratio in the Solar-type hot corinos is at least ten times larger than in the IRAS 4A outflows, again pointing to different grain mantles composition or densities. Interestingly, we noticed that CH$_3$OH/CH$_3$CHO tentatively decreases with increasing bolometric luminosity of the Class 0 hot corino; a possible reason could be the larger sizes of the hot corinos.
\end{itemize}

\begin{acknowledgements}
    We are very grateful to all the IRAM staff, whose dedication allowed us to carry out the SOLIS project. 
    This project has received funding from:
    (i) the European Research Council (ERC) under the European Union's Horizon 2020 research and innovation programme, for the Project “The Dawn of Organic Chemistry” (DOC), grant agreement No 741002;
    (ii) This work has been supported by the project PRIN-INAF 2016 The Cradle of Life - GENESIS-SKA (General Conditions in Early Planetary Systems for the rise of life with SKA);
    
    V.T. acknowledges the financial support from the European Union's Horizon 2020 research and innovation programme under the Marie Sklodowska-Curie grant agreement n. 664931.
    C.F. acknowledges support from the French National Research Agency in the framework of the Investissements d’Avenir program (ANR-15- IDEX-02), through the funding of the "Origin of Life" project of the Univ. Grenoble-Alpes
\end{acknowledgements}


\bibliographystyle{aa}
\bibliography{IRAS4A_outflows_new}

\begin{thebibliography}{78}
\expandafter\ifx\csname natexlab\endcsname\relax\def\natexlab#1{#1}\fi

\bibitem[{Altwegg {et~al.}(2016)Altwegg, Balsiger, Bar-Nun, Berthelier, Bieler,
  Bochsler, Briois, Calmonte, Combi, Cottin, De~Keyser, Dhooghe, Fiethe,
  Fuselier, Gasc, Gombosi, Hansen, Haessig, Ja~ckel, Kopp, Korth, Le~Roy, Mall,
  Marty, Mousis, Owen, Reme, Rubin, Semon, Tzou, Waite, \&
  Wurz}]{altwegg_prebiotic_2016}
Altwegg, K., Balsiger, H., Bar-Nun, A., {et~al.} 2016, Science Advances, 2,
  e1600285

\bibitem[{Arce {et~al.}(2008)Arce, Santiago-García, Jørgensen, Tafalla, \&
  Bachiller}]{arce_complex_2008}
Arce, H.~G., Santiago-García, J., Jørgensen, J.~K., Tafalla, M., \&
  Bachiller, R. 2008, The Astrophysical Journal Letters, 681, L21

\bibitem[{Bachiller {et~al.}(1998)Bachiller, Guilloteau, Gueth, Tafalla,
  Dutrey, Codella, \& Castets}]{bachiller_molecular_1998}
Bachiller, R., Guilloteau, S., Gueth, F., {et~al.} 1998, Astronomy and
  Astrophysics, 339, L49

\bibitem[{Balucani {et~al.}(2015)Balucani, Ceccarelli, \&
  Taquet}]{balucani_formation_2015}
Balucani, N., Ceccarelli, C., \& Taquet, V. 2015, Monthly Notices of the Royal
  Astronomical Society, 449, L16

\bibitem[{{Bennett} {et~al.}(2005{\natexlab{a}}){Bennett}, {Jamieson},
  {Osamura}, \& {Kaiser}}]{bennett_combined_2005}
{Bennett}, C.~J., {Jamieson}, C.~S., {Osamura}, Y., \& {Kaiser}, R.~I.
  2005{\natexlab{a}}, \apj, 624, 1097

\bibitem[{{Bennett} {et~al.}(2005{\natexlab{b}}){Bennett}, {Osamura}, {Lebar},
  \& {Kaiser}}]{bennett_laboratory_2005}
{Bennett}, C.~J., {Osamura}, Y., {Lebar}, M.~D., \& {Kaiser}, R.~I.
  2005{\natexlab{b}}, \apj, 634, 698

\bibitem[{Bianchi {et~al.}(2017)Bianchi, Codella, Ceccarelli, Taquet, Cabrit,
  Bacciotti, Bachiller, Chapillon, Gueth, Gusdorf, Lefloch, Leurini, Podio,
  Rygl, Tabone, \& Tafalla}]{bianchi_deuterated_2017}
Bianchi, E., Codella, C., Ceccarelli, C., {et~al.} 2017, Astronomy and
  Astrophysics, 606, L7

\bibitem[{Bianchi {et~al.}(2019)Bianchi, Codella, Ceccarelli, Vazart,
  Bachiller, Balucani, Bouvier, De~Simone, Enrique-Romero, Kahane, Lefloch,
  López-Sepulcre, Ospina-Zamudio, Podio, \& Taquet}]{bianchi_census_2019}
Bianchi, E., Codella, C., Ceccarelli, C., {et~al.} 2019, Monthly Notices of the
  Royal Astronomical Society, 483, 1850

\bibitem[{Blake {et~al.}(1995)Blake, Sandell, van Dishoeck, Groesbeck, Mundy,
  \& Aspin}]{blake_molecular_1995}
Blake, G.~A., Sandell, G., van Dishoeck, E.~F., {et~al.} 1995, The
  Astrophysical Journal, 441, 689

\bibitem[{Boogert {et~al.}(2015)Boogert, Gerakines, \&
  Whittet}]{boogert_observations_2015}
Boogert, A. C.~A., Gerakines, P.~A., \& Whittet, D. C.~B. 2015, Annual Review
  of Astronomy and Astrophysics, 53, 541

\bibitem[{Bottinelli {et~al.}(2004)Bottinelli, Ceccarelli, Lefloch, Williams,
  Castets, Caux, Cazaux, Maret, Parise, \& Tielens}]{bottinelli_complex_2004}
Bottinelli, S., Ceccarelli, C., Lefloch, B., {et~al.} 2004, The Astrophysical
  Journal, 615, 354

\bibitem[{Castets {et~al.}(2001)Castets, Ceccarelli, Loinard, Caux, \&
  Lefloch}]{castets_multiple_2001}
Castets, A., Ceccarelli, C., Loinard, L., Caux, E., \& Lefloch, B. 2001,
  Astronomy and Astrophysics, 375, 40

\bibitem[{Cazaux {et~al.}(2003)Cazaux, Tielens, Ceccarelli, Castets, Wakelam,
  Caux, Parise, \& Teyssier}]{cazaux_hot_2003}
Cazaux, S., Tielens, A. G. G.~M., Ceccarelli, C., {et~al.} 2003, The
  Astrophysical Journal, 593, L51

\bibitem[{Ceccarelli {et~al.}(2017)Ceccarelli, Caselli, Fontani, Neri,
  López-Sepulcre, Codella, Feng, Jiménez-Serra, Lefloch, Pineda, Vastel,
  Alves, Bachiller, Balucani, Bianchi, Bizzocchi, Bottinelli, Caux,
  Chacón-Tanarro, Choudhury, Coutens, Dulieu, Favre, Hily-Blant, Holdship,
  Kahane, Jaber Al-Edhari, Laas, Ospina, Oya, Podio, Pon, Punanova, Quenard,
  Rimola, Sakai, Sims, Spezzano, Taquet, Testi, Theulé, Ugliengo, Vasyunin,
  Viti, Wiesenfeld, \& Yamamoto}]{ceccarelli_seeds_2017}
Ceccarelli, C., Caselli, P., Fontani, F., {et~al.} 2017, The Astrophysical
  Journal, 850, 176

\bibitem[{Ceccarelli {et~al.}(2007)Ceccarelli, Caselli, Herbst, Tielens, \&
  Caux}]{ceccarelli_extreme_2007}
Ceccarelli, C., Caselli, P., Herbst, E., Tielens, A. G. G.~M., \& Caux, E.
  2007, Protostars and Planets V, 47

\bibitem[{Ceccarelli {et~al.}(1998)Ceccarelli, Castets, Loinard, Caux, \&
  Tielens}]{ceccarelli_detection_1998}
Ceccarelli, C., Castets, A., Loinard, L., Caux, E., \& Tielens, A. G. G.~M.
  1998, Astronomy and Astrophysics, 338, L43

\bibitem[{Charnley(2004)}]{charnley_acetaldehyde_2004}
Charnley, S.~B. 2004, Advances in Space Research, 33, 23

\bibitem[{Charnley {et~al.}(1992)Charnley, Tielens, \&
  Millar}]{charnley_molecular_1992}
Charnley, S.~B., Tielens, A. G. G.~M., \& Millar, T.~J. 1992, The Astrophysical
  Journal, 399, L71

\bibitem[{Choi(2001)}]{choi_high-resolution_2001}
Choi, M. 2001, The Astrophysical Journal, 553, 219

\bibitem[{Choi(2005)}]{choi_variability_2005}
Choi, M. 2005, The Astrophysical Journal, 630, 976

\bibitem[{Choi {et~al.}(2011)Choi, Kang, Tatematsu, Lee, \&
  Park}]{choi_radio_2011}
Choi, M., Kang, M., Tatematsu, K., Lee, J.-E., \& Park, G. 2011, Publications
  of the Astronomical Society of Japan, 63, 1281

\bibitem[{Codella {et~al.}(2019)Codella, Ceccarelli, Bianchi, Balucani, Podio,
  Caselli, Feng, Lefloch, López-Sepulcre, Neri, Spezzano, \&
  De~Simone}]{codella_seeds_2019}
Codella, C., Ceccarelli, C., Bianchi, E., {et~al.} 2019, Astronomy \&
  Astrophysics, submitted

\bibitem[{Codella {et~al.}(2017)Codella, Ceccarelli, Caselli, Balucani, Barone,
  Fontani, Lefloch, Podio, Viti, Feng, Bachiller, Bianchi, Dulieu,
  Jiménez-Serra, Holdship, Neri, Pineda, Pon, Sims, Spezzano, Vasyunin, Alves,
  Bizzocchi, Bottinelli, Caux, Chacón-Tanarro, Choudhury, Coutens, Favre,
  Hily-Blant, Kahane, Jaber Al-Edhari, Laas, López-Sepulcre, Ospina, Oya,
  Punanova, Puzzarini, Quenard, Rimola, Sakai, Skouteris, Taquet, Testi,
  Theulé, Ugliengo, Vastel, Vazart, Wiesenfeld, \&
  Yamamoto}]{codella_seeds_2017}
Codella, C., Ceccarelli, C., Caselli, P., {et~al.} 2017, Astronomy and
  Astrophysics, 605, L3

\bibitem[{Codella {et~al.}(2015)Codella, Fontani, Ceccarelli, Podio, Viti,
  Bachiller, Benedettini, \& Lefloch}]{codella_astrochemistry_2015}
Codella, C., Fontani, F., Ceccarelli, C., {et~al.} 2015, Monthly Notices of the
  Royal Astronomical Society, 449, L11

\bibitem[{Coutens {et~al.}(2016)Coutens, Jørgensen, van~der Wiel, Müller,
  Lykke, Bjerkeli, Bourke, Calcutt, Drozdovskaya, Favre, Fayolle, Garrod,
  Jacobsen, Ligterink, Öberg, Persson, van Dishoeck, \&
  Wampfler}]{coutens_alma-pils_2016}
Coutens, A., Jørgensen, J.~K., van~der Wiel, M. H.~D., {et~al.} 2016,
  Astronomy and Astrophysics, 590, L6

\bibitem[{De~Simone {et~al.}(2017)De~Simone, Codella, Testi, Belloche, Maury,
  Anderl, André, Maret, \& Podio}]{de_simone_glycolaldehyde_2017}
De~Simone, M., Codella, C., Testi, L., {et~al.} 2017, Astronomy and
  Astrophysics, 599, A121

\bibitem[{Di~Francesco {et~al.}(2001)Di~Francesco, Myers, Wilner, Ohashi, \&
  Mardones}]{di_francesco_infall_2001}
Di~Francesco, J., Myers, P.~C., Wilner, D.~J., Ohashi, N., \& Mardones, D.
  2001, The Astrophysical Journal, 562, 770

\bibitem[{Elsila {et~al.}(2009)Elsila, Glavin, \&
  Dworkin}]{elsila_cometary_2009}
Elsila, J.~E., Glavin, D.~P., \& Dworkin, J.~P. 2009, Meteoritics and Planetary
  Science, 44, 1323

\bibitem[{Enrique-Romero {et~al.}(2016)Enrique-Romero, Rimola, Ceccarelli, \&
  Balucani}]{enrique-romero_impossible?_2016}
Enrique-Romero, J., Rimola, A., Ceccarelli, C., \& Balucani, N. 2016, Monthly
  Notices of the Royal Astronomical Society, 459, L6

\bibitem[{Enrique-Romero {et~al.}(2019)Enrique-Romero, Rimola, Ceccarelli,
  Ugliengo, Balucani, \& Skouteris}]{enrique-romero_reactivity_2019}
Enrique-Romero, J., Rimola, A., Ceccarelli, C., {et~al.} 2019, ACS Earth and
  Space Chemistry, 3, 2158

\bibitem[{Garrod(2008)}]{garrod_new_2008}
Garrod, R.~T. 2008, Astronomy and Astrophysics, 491, 239

\bibitem[{Garrod \& Herbst(2006)}]{garrod_formation_2006}
Garrod, R.~T. \& Herbst, E. 2006, Astronomy and Astrophysics, 457, 927

\bibitem[{Goldsmith \& Langer(1999)}]{goldsmith_population_1999}
Goldsmith, P.~F. \& Langer, W.~D. 1999, The Astrophysical Journal, 517, 209

\bibitem[{Gueth {et~al.}(1996)Gueth, Guilloteau, \&
  Bachiller}]{gueth_precessing_1996}
Gueth, F., Guilloteau, S., \& Bachiller, R. 1996, Astronomy and Astrophysics,
  307, 891

\bibitem[{Herbst(2017)}]{herbst_synthesis_2017}
Herbst, E. 2017, International Reviews in Physical Chemistry, 36, 287

\bibitem[{Herbst \& van Dishoeck(2009)}]{herbst_complex_2009}
Herbst, E. \& van Dishoeck, E.~F. 2009, Annual Review of Astronomy and
  Astrophysics, 47, 427

\bibitem[{Holdship {et~al.}(2019)Holdship, Viti, Codella, Rawlings,
  Jimenez-Serra, Ayalew, Curtis, Habib, Lawrence, Warsame, \&
  Horn}]{holdship_observations_2019}
Holdship, J., Viti, S., Codella, C., {et~al.} 2019, The Astrophysical Journal,
  880, 138

\bibitem[{{Jacobsen, Steffen K.} {et~al.}(2019){Jacobsen, Steffen K.},
  {J\o{}rgensen, Jes K.}, {Di Francesco, James}, {Evans, Neal J.}, {Choi,
  Minho}, \& {Lee, Jeong-Eun}}]{jacobsen_organic_2018}
{Jacobsen, Steffen K.}, {J\o{}rgensen, Jes K.}, {Di Francesco, James}, {et~al.}
  2019, A\&A, 629, A29

\bibitem[{Jørgensen {et~al.}(2018)Jørgensen, Müller, Calcutt, Coutens,
  Drozdovskaya, Öberg, Persson, Taquet, van Dishoeck, \&
  Wampfler}]{jorgensen_alma-pils_2018}
Jørgensen, J.~K., Müller, H. S.~P., Calcutt, H., {et~al.} 2018, Astronomy and
  Astrophysics, 620, A170

\bibitem[{Jørgensen {et~al.}(2016)Jørgensen, van~der Wiel, Coutens, Lykke,
  Müller, van Dishoeck, Calcutt, Bjerkeli, Bourke, Drozdovskaya, Favre,
  Fayolle, Garrod, Jacobsen, Öberg, Persson, \&
  Wampfler}]{jorgensen_alma_2016}
Jørgensen, J.~K., van~der Wiel, M. H.~D., Coutens, A., {et~al.} 2016,
  Astronomy and Astrophysics, 595, A117

\bibitem[{Karska {et~al.}(2013)Karska, Herczeg, van Dishoeck, Wampfler,
  Kristensen, Goicoechea, Visser, Nisini, San José-García, Bruderer, Śniady,
  Doty, Fedele, Yıldız, Benz, Bergin, Caselli, Herpin, Hogerheijde,
  Johnstone, Jørgensen, Liseau, Tafalla, van~der Tak, \&
  Wyrowski}]{karska_water_2013}
Karska, A., Herczeg, G.~J., van Dishoeck, E.~F., {et~al.} 2013, Astronomy and
  Astrophysics, 552, A141

\bibitem[{{Kirchhoff} {et~al.}(1973){Kirchhoff}, {Johnson}, \&
  {Lovas}}]{kirchhoff_microwave_1973}
{Kirchhoff}, W.~H., {Johnson}, D.~R., \& {Lovas}, F.~J. 1973, Journal of
  Physical and Chemical Reference Data, 2, 1

\bibitem[{{Kleiner} {et~al.}(1996){Kleiner}, {Lovas}, \&
  {Godefroid}}]{kleiner_microwave_1996}
{Kleiner}, I., {Lovas}, F.~J., \& {Godefroid}, M. 1996, Journal of Physical and
  Chemical Reference Data, 25, 1113

\bibitem[{Kristensen {et~al.}(2012)Kristensen, van Dishoeck, Bergin, Visser,
  Yıldız, San Jose-Garcia, Jørgensen, Herczeg, Johnstone, Wampfler, Benz,
  Bruderer, Cabrit, Caselli, Doty, Harsono, Herpin, Hogerheijde, Karska, van
  Kempen, Liseau, Nisini, Tafalla, van~der Tak, \&
  Wyrowski}]{kristensen_water_2012}
Kristensen, L.~E., van Dishoeck, E.~F., Bergin, E.~A., {et~al.} 2012, Astronomy
  and Astrophysics, 542, A8

\bibitem[{{Lay} {et~al.}(1995){Lay}, {Carlstrom}, \& {Hills}}]{lay_ngc_1995}
{Lay}, O.~P., {Carlstrom}, J.~E., \& {Hills}, R.~E. 1995, \apjl, 452, L73

\bibitem[{Lefloch {et~al.}(1998)Lefloch, Castets, Cernicharo, Langer, \&
  Zylka}]{lefloch_cores_1998}
Lefloch, B., Castets, A., Cernicharo, J., Langer, W.~D., \& Zylka, R. 1998,
  Astronomy and Astrophysics, 334, 269

\bibitem[{Lefloch {et~al.}(2017)Lefloch, Ceccarelli, Codella, Favre, Podio,
  Vastel, Viti, \& Bachiller}]{lefloch_l1157-b1_2017}
Lefloch, B., Ceccarelli, C., Codella, C., {et~al.} 2017, Monthly Notices of the
  Royal Astronomical Society, 469, L73

\bibitem[{Loison {et~al.}(2014)Loison, Wakelam, \&
  Hickson}]{loison_interstellar_2014}
Loison, J.-C., Wakelam, V., \& Hickson, K.~M. 2014, Monthly Notices of the
  Royal Astronomical Society, 443, 398

\bibitem[{Looney {et~al.}(2000)Looney, Mundy, \& Welch}]{looney_unveiling_2000}
Looney, L.~W., Mundy, L.~G., \& Welch, W.~J. 2000, The Astrophysical Journal,
  529, 477

\bibitem[{López-Sepulcre {et~al.}(2017)López-Sepulcre, Sakai, Neri, Imai,
  Oya, Ceccarelli, Higuchi, Aikawa, Bottinelli, Caux, Hirota, Kahane, Lefloch,
  Vastel, Watanabe, \& Yamamoto}]{lopez-sepulcre_complex_2017}
López-Sepulcre, A., Sakai, N., Neri, R., {et~al.} 2017, Astronomy and
  Astrophysics, 606, A121

\bibitem[{Marcelino {et~al.}(2018)Marcelino, Gerin, Cernicharo, Fuente,
  Wootten, Chapillon, Pety, Lis, Roueff, Commerçon, \&
  Ciardi}]{marcelino_alma_2018}
Marcelino, N., Gerin, M., Cernicharo, J., {et~al.} 2018, Astronomy and
  Astrophysics, 620, A80

\bibitem[{Maury {et~al.}(2019)Maury, André, Testi, Maret, Belloche,
  Hennebelle, Cabrit, Codella, Gueth, Podio, Anderl, Bacmann, Bontemps, Gaudel,
  Ladjelate, Lefèvre, Tabone, \& Lefloch}]{maury_characterizing_2019}
Maury, A.~J., André, P., Testi, L., {et~al.} 2019, Astronomy and Astrophysics,
  621, A76

\bibitem[{Mendoza {et~al.}(2014)Mendoza, Lefloch, López-Sepulcre, Ceccarelli,
  Codella, Boechat-Roberty, \& Bachiller}]{mendoza_molecules_2014}
Mendoza, E., Lefloch, B., López-Sepulcre, A., {et~al.} 2014, Monthly Notices
  of the Royal Astronomical Society, 445, 151

\bibitem[{Millar {et~al.}(1991)Millar, Herbst, \&
  Charnley}]{millar_formation_1991}
Millar, T.~J., Herbst, E., \& Charnley, S.~B. 1991, The Astrophysical Journal,
  369, 147

\bibitem[{Müller {et~al.}(2005)Müller, Schlöder, Stutzki, \&
  Winnewisser}]{muller_cologne_2005}
Müller, H. S.~P., Schlöder, F., Stutzki, J., \& Winnewisser, G. 2005, Journal
  of Molecular Structure, 742, 215

\bibitem[{Neustock {et~al.}(1990)Neustock, Guarnieri, Demaison, \&
  Wlodarczak}]{neustock_millimeter_1990}
Neustock, W., Guarnieri, A., Demaison, J., \& Wlodarczak, G. 1990, Zeitschrift
  f{\"u}r Naturforschung A, 45, 702

\bibitem[{{{\"O}berg} {et~al.}(2009){{\"O}berg}, {Garrod}, {van Dishoeck}, \&
  {Linnartz}}]{oberg_formation_2009}
{{\"O}berg}, K.~I., {Garrod}, R.~T., {van Dishoeck}, E.~F., \& {Linnartz}, H.
  2009, \aap, 504, 891

\bibitem[{Ospina-Zamudio {et~al.}(2018)Ospina-Zamudio, Lefloch, Ceccarelli,
  Kahane, Favre, López-Sepulcre, \& Montarges}]{ospina-zamudio_first_2018}
Ospina-Zamudio, J., Lefloch, B., Ceccarelli, C., {et~al.} 2018, Astronomy and
  Astrophysics, 618, A145

\bibitem[{Ospina-Zamudio {et~al.}(2019)Ospina-Zamudio, Lefloch, Favre,
  López-Sepulcre, Bianchi, Ceccarelli, De~Simone, Bouvier, \&
  Kahane}]{ospina-zamudio_molecules_2019}
Ospina-Zamudio, J., Lefloch, B., Favre, C., {et~al.} 2019, Monthly Notices of
  the Royal Astronomical Society, 490, 2679

\bibitem[{Padovani {et~al.}(2016)Padovani, Marcowith, Hennebelle, \&
  Ferrière}]{padovani_protostars:_2016}
Padovani, M., Marcowith, A., Hennebelle, P., \& Ferrière, K. 2016, Astronomy
  and Astrophysics, 590, A8

\bibitem[{Pickett {et~al.}(1998)Pickett, Poynter, Cohen, Delitsky, Pearson, \&
  Müller}]{pickett_submillimeter_1998}
Pickett, H.~M., Poynter, R.~L., Cohen, E.~A., {et~al.} 1998, Journal of
  Quantitative Spectroscopy and Radiative Transfer, 60, 883

\bibitem[{Pizzarello {et~al.}(2006)Pizzarello, Cooper, \&
  Flynn}]{pizzarello_nature_2006}
Pizzarello, S., Cooper, G.~W., \& Flynn, G.~J. 2006, Meteorites and the Early
  Solar System II, 625

\bibitem[{Podio {et~al.}(2016)Podio, Codella, Gueth, Cabrit, Maury, Tabone,
  Lefèvre, Anderl, André, Belloche, Bontemps, Hennebelle, Lefloch, Maret, \&
  Testi}]{podio_first_2016}
Podio, L., Codella, C., Gueth, F., {et~al.} 2016, Astronomy and Astrophysics,
  593, L4

\bibitem[{{Podio} {et~al.}(2014){Podio}, {Lefloch}, {Ceccarelli}, {Codella}, \&
  {Bachiller}}]{podio_molecular_2014}
{Podio}, L., {Lefloch}, B., {Ceccarelli}, C., {Codella}, C., \& {Bachiller}, R.
  2014, \aap, 565, A64

\bibitem[{Santangelo {et~al.}(2015)Santangelo, Codella, Cabrit, Maury, Gueth,
  Maret, Lefloch, Belloche, André, Hennebelle, Anderl, Podio, \&
  Testi}]{santangelo_jet_2015}
Santangelo, G., Codella, C., Cabrit, S., {et~al.} 2015, Astronomy and
  Astrophysics, 584, A126

\bibitem[{Skouteris {et~al.}(2018)Skouteris, Balucani, Ceccarelli, Vazart,
  Puzzarini, Barone, Codella, \& Lefloch}]{skouteris_genealogical_2018}
Skouteris, D., Balucani, N., Ceccarelli, C., {et~al.} 2018, The Astrophysical
  Journal, 854, 135

\bibitem[{Skouteris {et~al.}(2017)Skouteris, Vazart, Ceccarelli, Balucani,
  Puzzarini, \& Barone}]{skouteris_new_2017}
Skouteris, D., Vazart, F., Ceccarelli, C., {et~al.} 2017, Monthly Notices of
  the Royal Astronomical Society, 468, L1

\bibitem[{Smith {et~al.}(2000)Smith, Bonnell, Emerson, \&
  Jenness}]{smith_ngc_2000}
Smith, K.~W., Bonnell, I.~A., Emerson, J.~P., \& Jenness, T. 2000, Monthly
  Notices of the Royal Astronomical Society, 319, 991

\bibitem[{Taquet {et~al.}(2012{\natexlab{a}})Taquet, Ceccarelli, \&
  Kahane}]{taquet_multilayer_2012}
Taquet, V., Ceccarelli, C., \& Kahane, C. 2012{\natexlab{a}}, Astronomy and
  Astrophysics, 538, A42

\bibitem[{Taquet {et~al.}(2012{\natexlab{b}})Taquet, Ceccarelli, \&
  Kahane}]{taquet_formaldehyde_2012}
Taquet, V., Ceccarelli, C., \& Kahane, C. 2012{\natexlab{b}}, The Astrophysical
  Journal, 748, L3

\bibitem[{Taquet {et~al.}(2019)Taquet, Codella, DeSimone, López-Sepulcre,
  Pineda, Segura-Cox~D., Ceccarelli, Caselli, Cabrit, Gusdorf, pineau~des
  forêts, persson, Alves, Caux, Favre, Fontani, Neri, Oya, Sakai, Vastel, \&
  Yamamoto}]{taquet_seeds_2019}
Taquet, V., Codella, C., DeSimone, M., {et~al.} 2019, Astronomy \&
  Astrophysics, submitted

\bibitem[{Taquet {et~al.}(2015)Taquet, López-Sepulcre, Ceccarelli, Neri,
  Kahane, \& Charnley}]{taquet_constraining_2015}
Taquet, V., López-Sepulcre, A., Ceccarelli, C., {et~al.} 2015, The
  Astrophysical Journal, 804, 81

\bibitem[{Tielens \& Hagen(1982)}]{tielens_model_1982}
Tielens, A. G. G.~M. \& Hagen, W. 1982, Astronomy and Astrophysics, 114, 245

\bibitem[{Turner(1990)}]{turner_detection_1990}
Turner, B.~E. 1990, The Astrophysical Journal, 362, L29

\bibitem[{Vasyunin {et~al.}(2017)Vasyunin, Caselli, Dulieu, \&
  Jiménez-Serra}]{vasyunin_formation_2017}
Vasyunin, A.~I., Caselli, P., Dulieu, F., \& Jiménez-Serra, I. 2017, The
  Astrophysical Journal, 842, 33

\bibitem[{Watanabe \& Kouchi(2002)}]{watanabe_efficient_2002}
Watanabe, N. \& Kouchi, A. 2002, The Astrophysical Journal, 571, L173

\bibitem[{Xu {et~al.}(2008)Xu, Fisher, Lees, Shi, Hougen, Pearson, Drouin,
  Blake, \& Braakman}]{xu_torsion_2008}
Xu, L.-H., Fisher, J., Lees, R., {et~al.} 2008, Journal of Molecular
  Spectroscopy, 251, 305

\bibitem[{Zucker {et~al.}(2018)Zucker, Schlafly, Speagle, Green, Portillo,
  Finkbeiner, \& Goodman}]{zucker_mapping_2018}
Zucker, C., Schlafly, E.~F., Speagle, J.~S., {et~al.} 2018, The Astrophysical
  Journal, 869, 83

\end{thebibliography}
  
\end{document}